\documentclass[nofootinbib,twocolumn,] {revtex4-1} 
\usepackage{graphicx}
\usepackage{upgreek}

\newcommand{\mymu}{{\ensuremath \upmu}}
\newcommand{\muz}{{\ensuremath \mu_\mathrm{0}}}

\newcommand{\smco}{Sm$_2$Co$_{17}$}

\begin{document}
\title{Construction, Measurement, Shimming, and Performance of the NIST-4 Magnet System}
\author{F. Seifert} 
\author{S. Li}
\author{B. Han}
\author{L. Chao}
\author{A. Cao}
\author{D. Haddad}
\affiliation{Fundamental Electrical Measurements Group, National Institute of Standards and Technology, Gaithersburg, MD 20899}
\author{H. Choi}
\author{L. Haley}  
\affiliation{Electron Energy Corporation, Landisville, PA 17538}
\author{S. Schlamminger} 
\affiliation{Fundamental Electrical Measurements Group, National Institute of Standards and Technology, Gaithersburg, MD 20899}


\date{\today}



\begin{abstract}
The magnet system is one of the key elements of a watt balance. For the new watt balance currently under construction at the National Institute of Standards and Technology, a permanent magnet system was chosen. We describe the detailed construction of the magnet system, first measurements of the field profile, and shimming techniques that were used to achieve a flat field profile. The relative change of the radial magnetic flux density is less than $10^{-4}$ over a range of 5\,cm. We further characterize the most important aspects of the magnet and give order of magnitude estimates for several systematic effects that originate from the magnet system.
\end{abstract}
\maketitle
\section{Introduction}
A redefinition of the International System of Units, the SI, is impending and might occur as early as 2018.  A system of seven reference constants will replace the seven base units that form the present foundation of our unit system~\cite{mills11}. Specifically in the context of mass metrology, the base unit kilogram will be replaced by a fixed value of the Planck constant. With this transition, the International Prototype of the Kilogram (IPK), will lose its status as being the only weight on Earth, whose mass is known with zero uncertainty. In the future, mass will be realized from a fixed value of the Planck constant by various means. A promising apparatus to realize mass at the kilogram level is the watt balance~\cite{steiner13,kibble75}. Watt balances have a long history at the National Institute of Standards and Technology (NIST). In 1980,  NIST's first watt balance was designed to realize the absolute ampere and then later to measure the Planck constant~\cite{Olsen80}. In the past three and a half decades, several measurements of the Planck constant have been published, the most recent in 2014~\cite{Schlamminger14}. Currently, a new watt balance, NIST-4, is being designed and built. This watt balance will be used to realize the unit of mass in the United States. 

A watt balance is a force transducer that can be calibrated in absolute terms using voltage, resistance, frequency, and length reference standards, i.e., without dead weights. The instrument is used in two modes, typically referred to as force mode and velocity mode. In force mode, the gravitational force of a mass, $mg$, is compensated by an electromagnetic force. The electromagnetic force is produced by a current in a coil that is immersed in a radial magnetic field. The equation governing the force mode is $mg = I Bl$, where $I$ is the current in the coil, $l$ the wire length of the coil, and $B$ the magnetic flux density of the field at the coil position. The local acceleration $g$ and the current $I$ can be measured using dedicated instruments. The only term that needs calibration is the flux integral $Bl$. This integral can be calibrated to very high precision in velocity mode. The coil is moved through the magnetic field with constant velocity $v$ yielding an induced voltage, $U=v Bl$. The flux integral is inferred by dividing the voltage by the velocity. By using this calibrated value of $Bl$ in the equation of the force mode, the value for the mass can be obtained by
\begin{equation}
m = \frac{U I}{g v}.
\end{equation}
The equation above connects mass to electrical quantities: current and voltage.  The electrical quantities can be linked to the Planck constant and two frequencies using the Josephson effect and the quantum Hall effect. This connection is beyond the scope of this article. A review can be found in~\cite{steiner13}.

The considerations that led to the design of the permanent magnet system described here are given in~\cite{ss13}. While the findings in~\cite{ss13} were based on simulations and theoretical calculations, this article presents measurements that were made on the real magnet system.  We describe in detail the construction of the magnet system and focus on the implications for the performance of NIST-4.

\section{The basic design}
\begin{figure}[h!]
\centering
\begin{minipage}[c]{2.3in}
\includegraphics[width=2.2in]{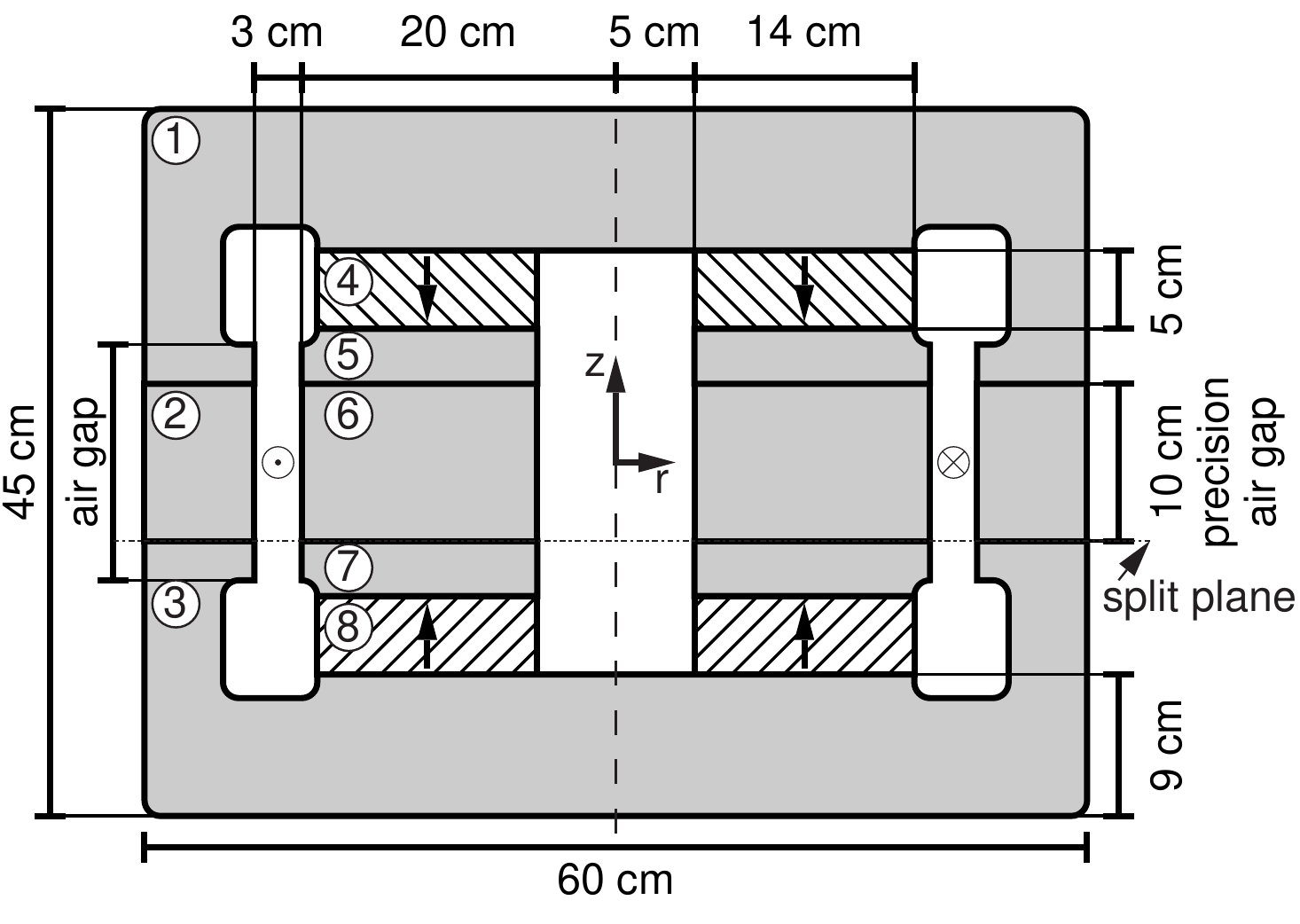}
\end{minipage}
\begin{minipage}[c]{1.1in}
\includegraphics[width=1.0in]{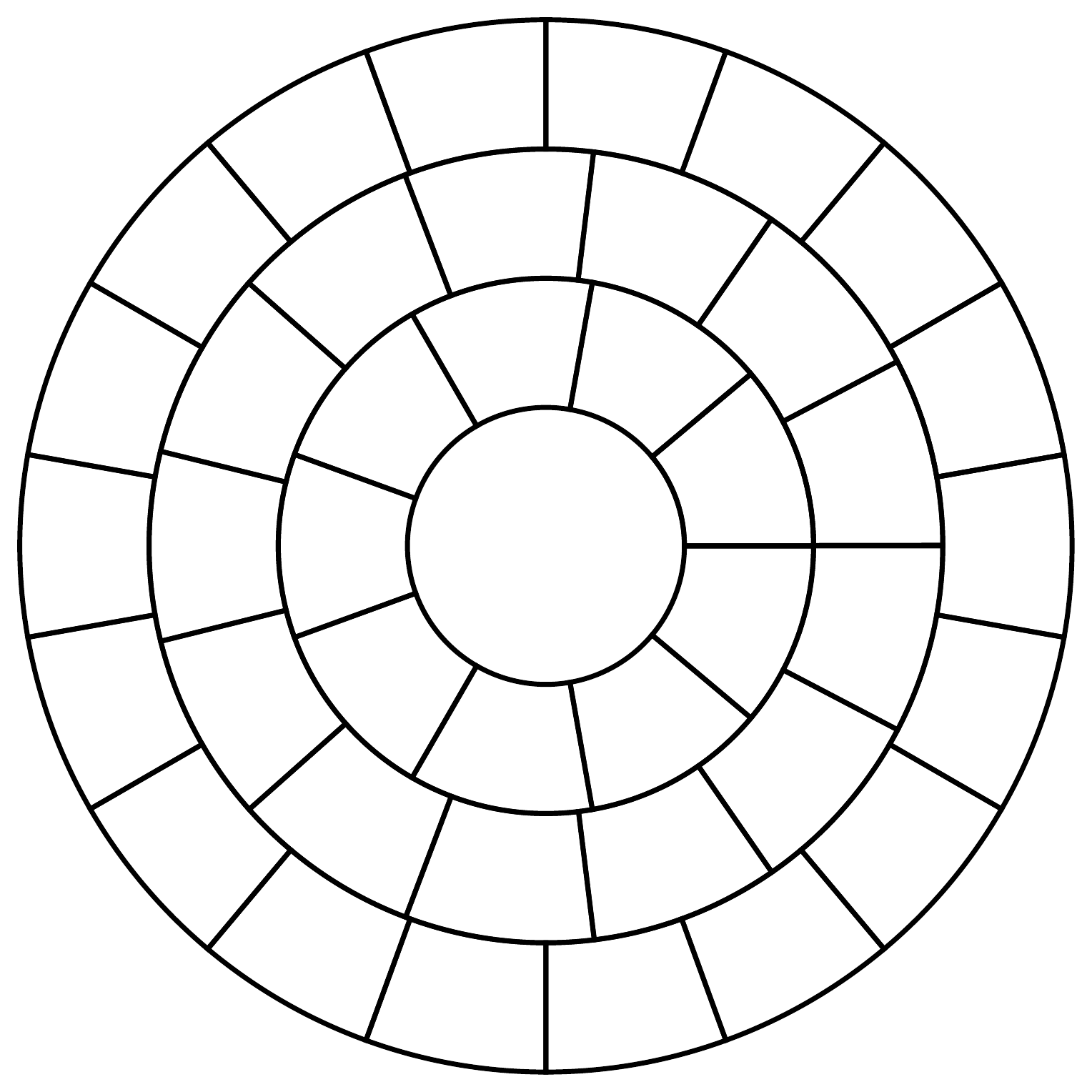}
\end{minipage}
\caption{Shown on the left is a cross sectional view of the magnet system. The assembly exhibits azimuthal and up-down symmetry. The gray parts are made from AISI 1021 steel and the hatched parts from \smco. The arrows indicate the direction of magnetization. A total of eight large components are required to build the magnet system. The components are indicated by the encircled numbers: 1=upper yoke cap, 2=outer yoke, 3=lower yoke cap, 4=upper \smco\ ring, 5=upper inner yoke, 6=middle inner yoke, 7=lower inner yoke, 8=lower \smco\ ring. The circles in the precision air gap indicate the current in the coil.  The segmentation of the \smco\ rings is shown on the right.}
\label{fig:circuitClose}
\end{figure}

The design of the NIST-4 magnet system was inspired by a magnet design put forward by the BIPM watt balance group~\cite{ms07}. In our design, shown in Fig.~\ref{fig:circuitClose}, two Sm$_2$Co$_{17}$ rings are opposing each other and their magnetic flux is guided by  low-carbon steel, also referred to as mild or soft steel, through a cylindrical air gap. The gap has a width of 3\,cm and is in total 15\,cm long. The inner 10\,cm of the gap is called the precision air gap  and it is desired to have a very uniform field in the central 8\,cm of this precision air gap. Short of twelve access holes in each the top and bottom, the gap is entirely enclosed by iron.

\begin{figure}[h!]
\centering
\begin{minipage}[c]{1.7in}
\includegraphics[scale=1.2,viewport = 6 6 105 105,clip]{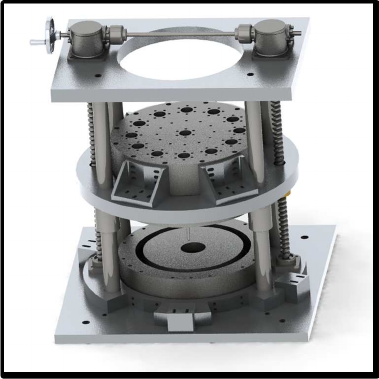} 
\end{minipage}
\begin{minipage}[c]{1.7in}
\includegraphics[scale=1.0,viewport = 1 1 133 129,clip]{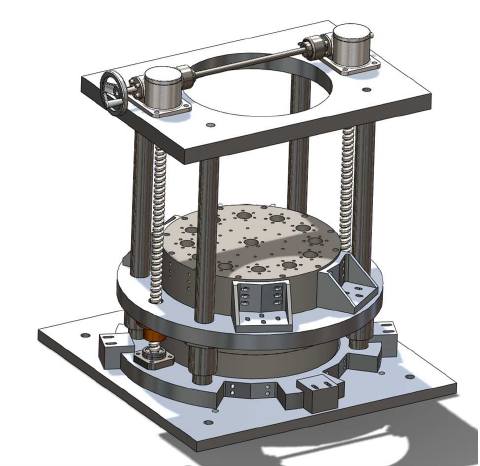} 
\end{minipage}
\caption{The picture on the left shows a rendering of the magnet in the magnet-splitter with the magnet in the open state. On the right is shown a technical drawing of the magnet in the magnet-splitter in the closed state. In order to install the magnet in the splitter, the magnet is craned onto the base plate. Then, the splitter can be slipped over the magnet using a crane. Finally, the middle ring is fastened to the magnet using four angle brackets in the middle and 16 bolts through the lower ring.}
\label{fig:splitter}
\end{figure}

In order to insert the coil into the air gap, the magnet can be split open such that the top two thirds of the magnet separate from the bottom third. CAD drawings of the magnet and the splitter are shown in Fig.~\ref{fig:splitter}.  A cross-sectional view of the basic design is shown in Fig.~\ref{fig:circuitClose}. The 8 basic components of the magnet are indicated by encircled numbers. We refer to components 2 and 6 as the outer yoke and inner yoke, respectively.

While NIST was responsible for the schematic design of the magnet, the detailed design and manufacturing was contracted to Electron Energy Corporation (EEC)\footnote{Certain commercial equipment, instruments, or materials are identified in this paper in order to specify the experimental procedure adequately. Such identification is not intended to imply recommendation or endorsement by the National Institute of Standards and Technology, nor is it intended to imply that the materials or equipment identified are necessarily the best available for the purpose.}. 
During the manufacturing process a few changes were made to improve the performance of the magnet. One such change pertains to the grade of the low-carbon steel used to produce the yoke parts. While in~\cite{ss13} the parts were identified to be made from A36, instead AISI 1021 steel was used to make the parts. This change was made because a large ingot of AISI 1021 could be purchased that allowed building all yoke parts from a single casting. By using raw material from one cast, a better homogeneity of the magnetic properties can be ensured in the final product. Both alloys are low-carbon steels, i.e., less than 0.3\% carbon by weight. The weight fraction of the carbon content of A36 steel is on average 0.05\% higher than that of AISI 1021. Other than the improved homogeneity of the material, this change is insignificant for the performance of the magnetic circuit.

\begin{figure}[h!]
\centering
\includegraphics[width=3.3in]{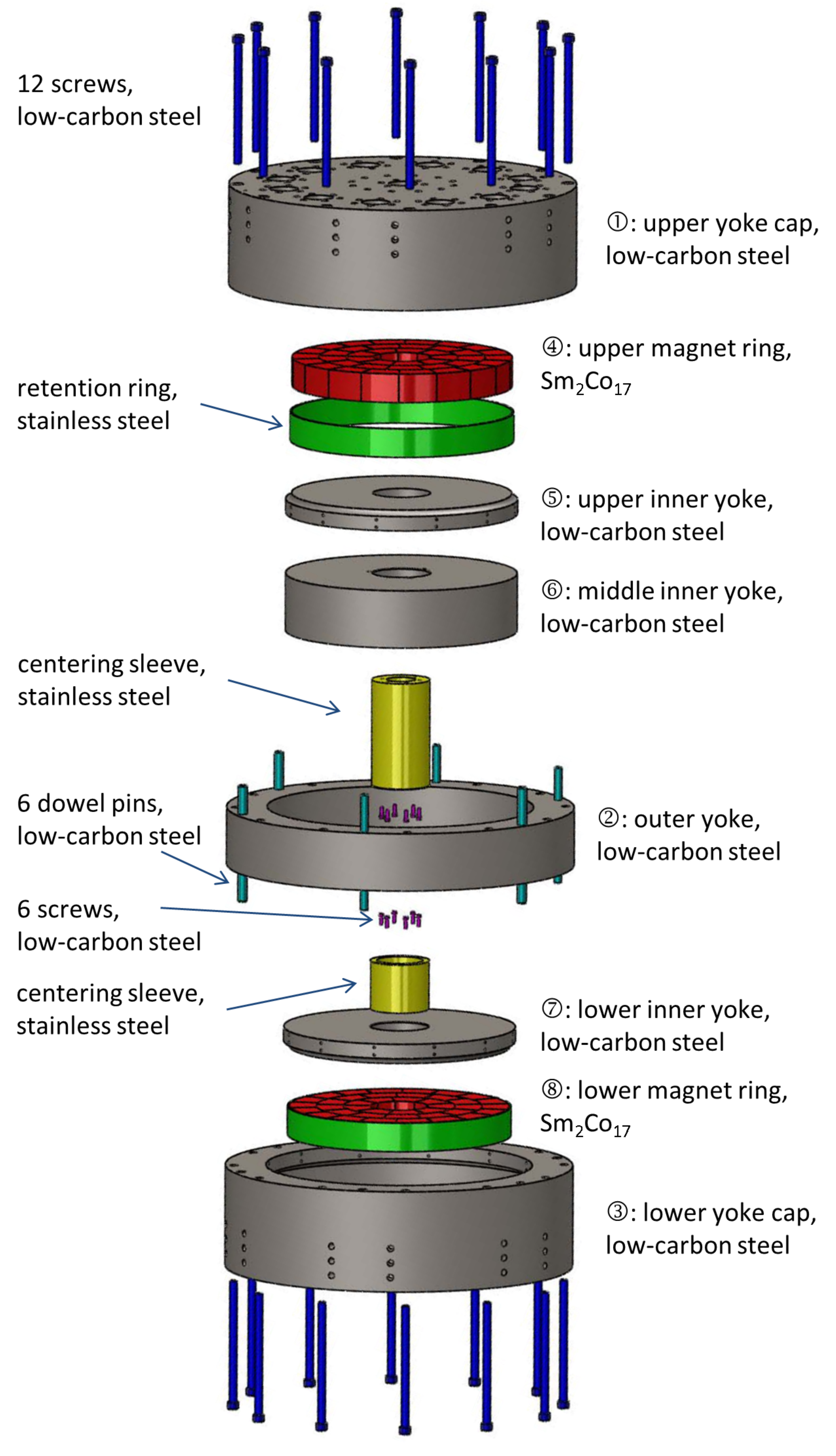}
\caption{Exploded view of the magnet. The eight major components shown in the cross sectional view in Fig.~\ref{fig:circuitClose} are labeled on the right, and additional hardware is labeled on the left. The circled numbers correspond to the numbers in the cross sectional drawing. The purpose of the centering sleeves is to center the inner yokes and the magnet rings on the yoke caps. Dowel pins ensure that the magnet only opens in the vertical direction. The stainless steel bands constrain the magnets in the radial direction.}
\label{fig:explodedview}
\end{figure}

In the final design, shown in Fig.~\ref{fig:explodedview}, two stainless steel sleeves were added to center the \smco\ and the inner yokes. Also, two stainless steel bands around the \smco\ magnet rings were added to aid the assembly process. In addition, dowel pins made from low-carbon steel allow us to reproducibly open and close the magnet.

\section{Material properties}
Three different materials were used in constructing the permanent magnet system: Sm$_2$Co$_{17}$, low-carbon steel 1021, and stainless steel. The stainless steel parts were annealed to reduce the relative magnetic permeability to near unity and are therefore irrelevant for the magnetic circuit. Hence, the stainless parts are not considered any further.

\subsection{Permanent Magnet -- \smco}
Two \smco\ rings, with a combined mass of 91\,kg, form the active magnetic material. Because it requires a lot of power and a large fixture to magnetize one ring, each ring was segmented in 40 pieces, see the sketch in Fig.~\ref{fig:circuitClose}. Each piece was individually magnetized. The segmentation was carried out in three concentric rings comprised of 9, 13, and 18 segments each. The largest segments in the outer ring has a volume of 138.6\,cm$^3$.  The \smco\ rings had to be assembled with the segments fully magnetized. To facilitate this assembly process and to keep the repulsive forces between individual segments under control, the rings were assembled on the inner yoke pieces using vacuum compatible epoxy. In addition, a stainless steel band around the ring, as shown in Fig.~\ref{fig:explodedview}, contains the \smco\ segments in the radial direction.

\begin{figure}[h!]
\centering
\includegraphics[width=2.7in]{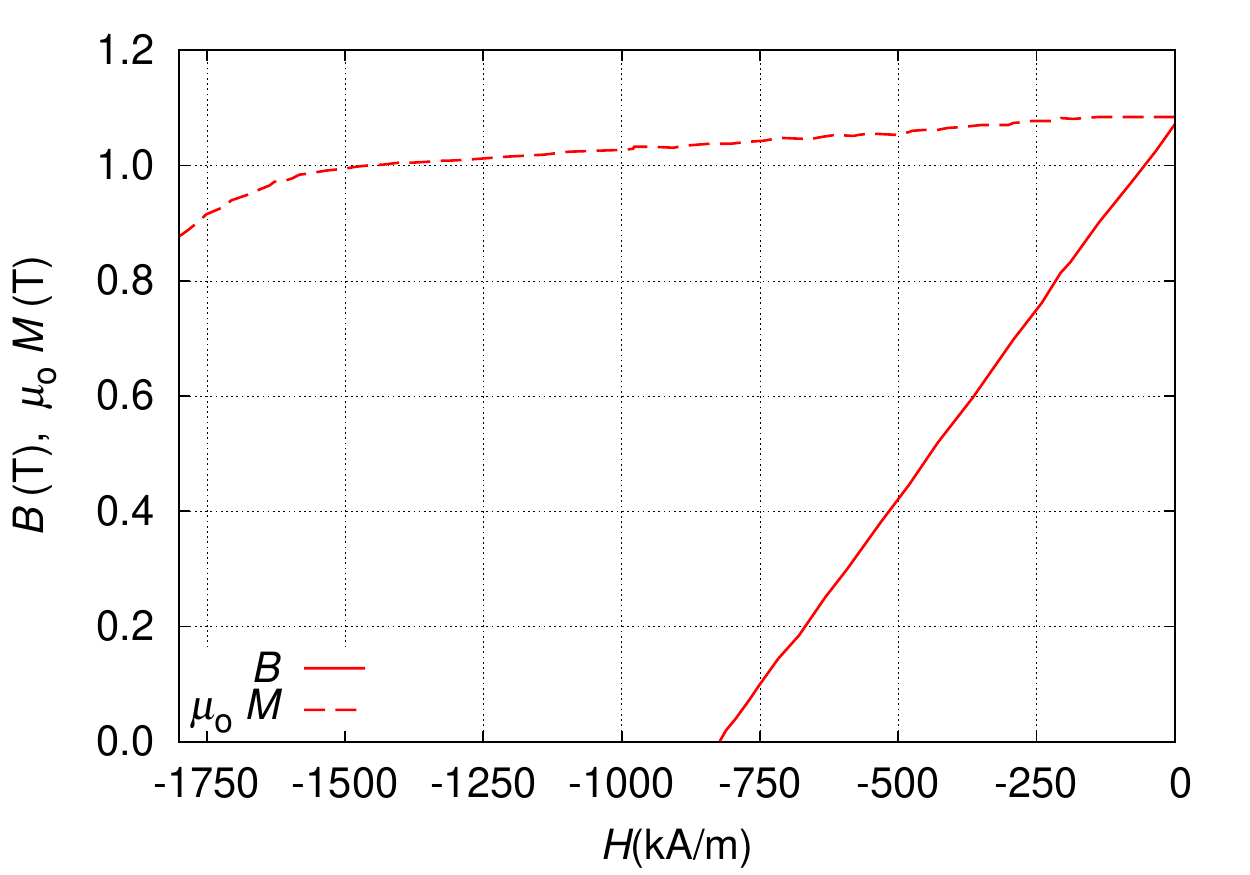}
\caption{Measured demagnetization curve on a Sm$_2$Co$_{17}$ sample. The sample was measured at 26\,$^\circ$C.}
\label{fig:SmCO:BH}
\end{figure}

In order to verify the magnetic properties of the \smco, five cylindrical test specimens (10\,mm diameter and 10\,mm height) were fabricated in addition to the 80 segments.
The magnetization curve of these samples were measured at EEC. Fig.~\ref{fig:SmCO:BH} shows the measurement of one such sample. This sample had a remanent flux density of 1.08\,T and a maximum energy product of $(BH)_{\mathrm{max}}=224.7$\,kJ/m$^3$. Of the five samples tested the remanence values were within 0.2\,\% and the maximum energy density within 0.6\,\% of each other. 

 For each of the 80 segments, the total flux was measured. After all measurements were obtained and recorded, a position for each segment was chosen to ensure uniform magnetization in the azimuthal direction and between the two rings. After assembly, the total flux values of the two ring magnet assemblies were measured and found to be within 0.2\,\% of each other.

\subsection{Yoke -- 1021 steel}
The yoke of the magnet is made from AISI 1021 carbon steel. To verify the composition, five samples were taken from the material and a chemical analysis was performed using AES (Atom Emission Spectroscopy). All samples conformed to the steel grade 1021. In the five samples, the carbon fraction varied from 0.20\,\% to 0.23\,\% and the manganese fraction from 0.87\,\% to 0.88\,\%. Phosphorus and sulfur had a relative weight of 0.013\,\% and 0.012\,\% respectively. The yoke parts were annealed after machining by heating to 850\,$^\circ$C for at least 4 hours followed by a slow cool down. The outside parts of the yoke were nickel coated to prevent corrosion. The inside parts and the surfaces that are relevant for the magnetic circuit were not nickel coated. Instead, the surfaces were coated with a small amount of vacuum compatible oil (Krytox 1506) to prevent oxidation.

The magnetic properties of the low-carbon steel were investigated using two toroidal samples made from the same ingot as the magnet yoke. After machining, one sample was annealed using the same recipe as the yoke parts, the other sample was not heat treated after machining. The results from the annealed samples are relevant for the NIST-4 magnet system. However, the results of the non annealed sample serve as a reference and worst case scenario. On each toroid, two sets of windings were placed: An excitation winding (1) and a pick-up winding (2). Each winding had $N_1=N_2=200$~turns. The toroidal cores had a rectangular cross section with inner and outer radii of $r_{\mathrm{i}}$ and $r_{\mathrm{o}}$, respectively. Each toroid had slightly different dimensions. The mean radius, $r_\mathrm{m}=(1/2) (r_\mathrm{i}+r_\mathrm{o})$, of the annealed sample was 38.0\,mm and that of the not annealed sample was 30.2\,mm. In the first case, the cross sectional area was $A=5.73\times 10^{-5}$\,m$^2$ and in the second case, $6.09\times 10^{-5}$\,m$^2$, respectively. Sinusoidal current with a frequency of 0.4\,Hz was sent through the excitation winding and the induced voltage, $V(t)$, was measured across the pick-up winding. The current in the excitation coil was measured as a voltage drop across a series $1\,\Omega$ resistor. The magnetic field $H_{\mathrm{b}}(t)$ generated by the current in the excitation winding is calculated using Ampere's law,
\begin{equation}
H_{\mathrm{b}}(t) = \frac{N_1i(t)}{2\pi r_{\mathrm{m}}}\label{eq:H_of_t}.
\end{equation}
The derivative of the total flux, which we assume to be uniformly distributed and normal to the cross section of the toroid, is
\begin{equation}
\frac{dB}{dt} = -\frac{V(t)}{N_2A} \label{eq:B_of_t}.
\end{equation}
The magnetic flux density is found by integrating (\ref{eq:B_of_t}), where the constant of integration is chosen such that $\int B(t)\mbox{d}t=0$ over one cycle.
The relative permeability $\mu_{\mathrm{r}}$ and differential permeability $\mu_{\mathrm{d}}$ of the yoke can be found as a function of the magnetizing field, from the hysteresis curves using
\begin{equation}
\mu_{\mathrm{r}} = \frac{1}{\muz}\frac{B}{H} \;\;\;\mbox{and}\;\;\;\mu_{\mathrm{d}} =  \frac{1}{\muz}\frac{\mbox{d}B}{\mbox{d}H}.
\end{equation}
Five sets of measurements were taken for each sample. After each set, the magnetized core was degaussed by subjecting it to a damped AC field, with an amplitude higher than $H_{\mathrm{sat}}$ and gradually reducing the amplitude to zero. Because of the low excitation frequency, the magnetic measurements can be considered as pseudo-static, allowing us to neglect eddy-current effects on these measurements. 

\begin{figure}[h!]
\centering
\includegraphics[width=3.5in]{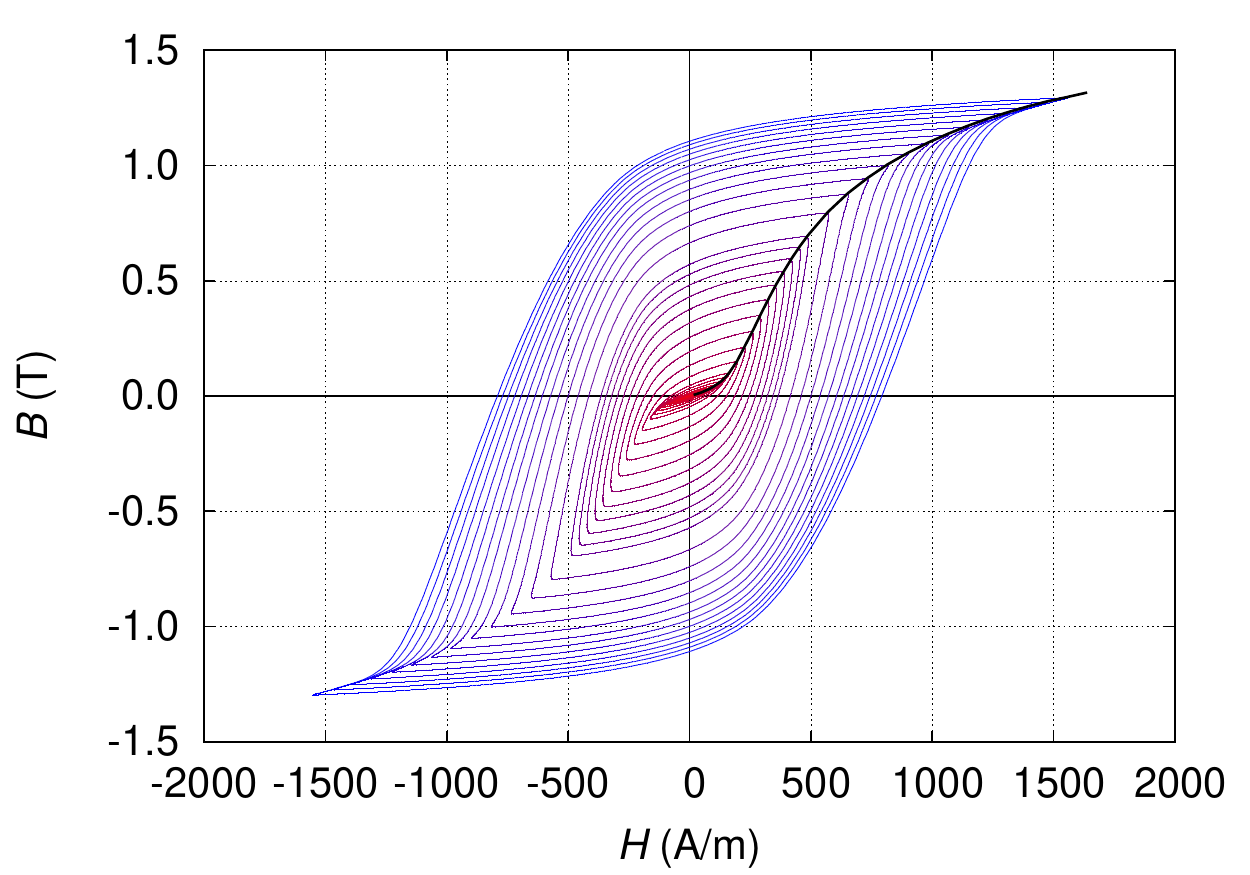}
\caption{One set of $B$-$H$ measurements for the annealed sample. A total of ten sets, five for the annealed sample and five for the non annealed sample were taken. The thick line is the normal hysteresis curve.}
\label{fig:BvsH}
\end{figure}

Fig.~\ref{fig:BvsH} shows a set of hysteresis curves with the normal hysteresis curve~\cite{Bozorth} for the annealed sample, obtained by progressively varying the amplitude of the AC excitation current.  The saturation field is derived from the magnetization curve ($M$-$H$) and is found to be $H_{\mathrm{sat}}=1.64$\,kA/m. 

\begin{figure}[h!]
\centering
\includegraphics[width=3.5in]{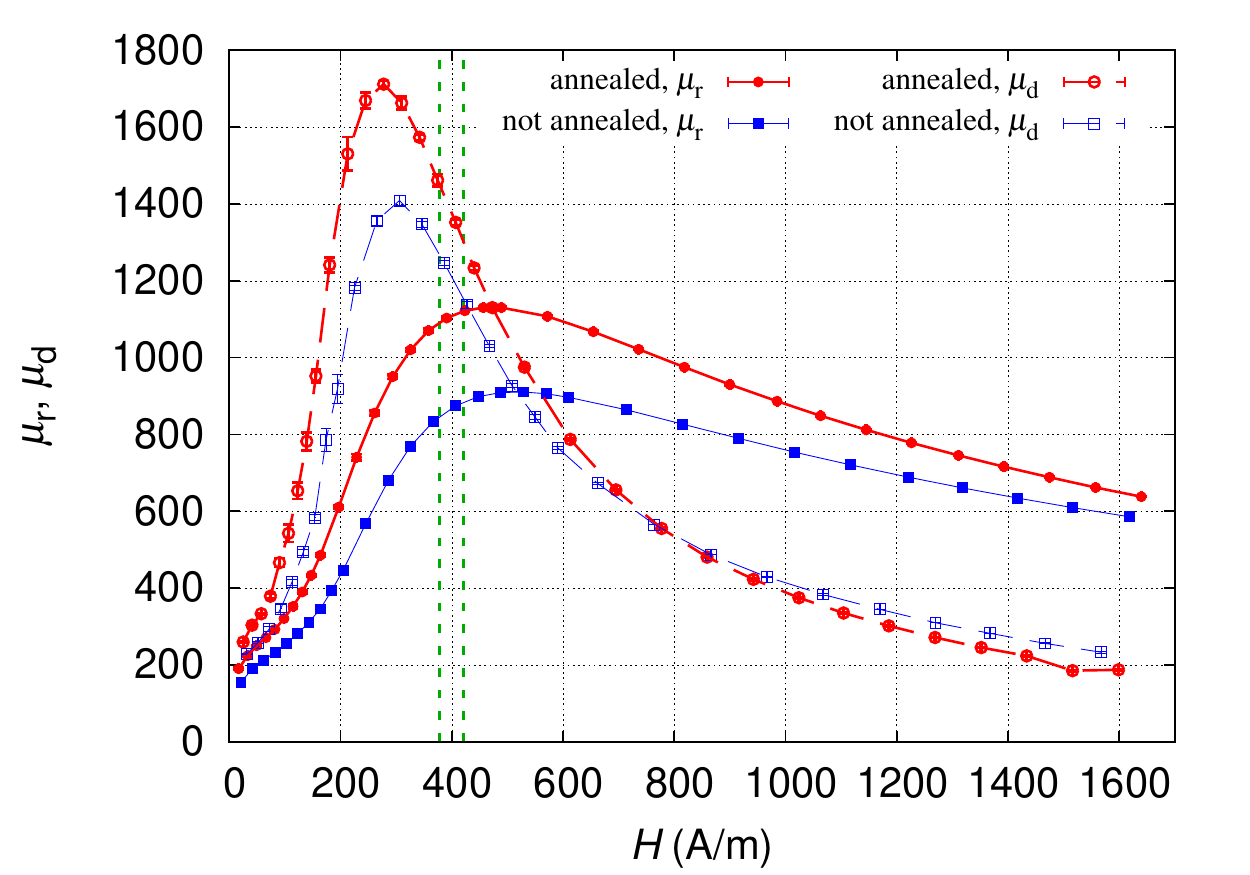}
\caption{The differential (dashed lines) and relative (solid lines) permeability of the annealed (circles) and non annealed sample (squares). The data points are obtained from five sets shown in Fig.~\ref{fig:BvsH}. The vertical dotted lines show the magnetic field at which the iron parts adjacent to the coil operated. The vertical error bars are the standard deviation of the five sets of measurements that were taken for each sample.}
\label{fig:perm}
\end{figure}

Fig.~\ref{fig:perm} shows the relative ($\mu_{\mathrm{r}}$) and differential ($\mu_{\mathrm{d}}$) permeability curves for the annealed and non annealed samples derived from the normal hysteresis curve. The point where the $\mu_{\mathrm{d}}$ and $\mu_{\mathrm{r}}$ curves intersect is the maximum relative permeability $\mu_{\mathrm{m}}$ of the yoke. Results indicate that annealing the yoke increases $\mu_{\mathrm{m}}$ by a factor $\approx1.2$.

The point at which the yoke operates in the $\mu_{\mathrm{r}}$-$H$ plot shown in Fig.~\ref{fig:perm} can be found by combining a measurement with the hysteresis data. In the center of the gap, $r_{\mathrm{c}}=215$\,mm and the magnetic flux density is $B_{\mathrm{c}}=0.55$\,T. Inside the gap, the magnetic field follows a $1/r$ relationship, $B(r)=B_{\mathrm{c}} r_{\mathrm{c}}/r$, hence the magnetic flux density at the surface of the inner/outer yoke can be calculated to be $B_{\mathrm{iy}}=0.59$\,T/$B_{\mathrm{oy}}=0.51$\,T, respectively. On the normal hysteresis curve (Fig.~\ref{fig:BvsH}), these values correspond to $H_{\mathrm{iy}}=420$\,A/m and $H_{\mathrm{oy}}=380$\,A/m, which are close to the maximum of $\mu_{\mathrm{r}}(H)$, yielding a value of $\mu_{\mathrm{r}} \approx 1100$.

Operating the yoke near the maximum value of $\mu_{\mathrm{r}}$, makes the reluctance of the yoke, to first order, independent of the field $H$. This is the preferred operating point for a watt balance, because the reluctance of the magnetic circuit is independent of the weighing current. In our magnet, we are not quite at  the maximum value of $\mu_{\mathrm{r}}$, but close. The effect of the weighing current on the yoke reluctance needs to be analyzed in detail. With the measurements shown in Figs.~\ref{fig:BvsH} and \ref{fig:perm}, we provide a basis to further model these effects.

\subsection{Temperature dependence of the magnetic flux density in the gap}
The temperature dependence of the radial magnetic flux density in the gap is governed primarily by the temperature coefficient of the \smco. In addition, the flux density depends, to a smaller extent, on  changes in reluctance of the magnetic circuit caused by temperature dependence of the permeability of the iron and changes in geometry due to thermal expansion. The temperature coefficient of the magnetic flux density in the gap was measured and found to be 
\begin{equation}
\frac{\Delta B_{\mathrm{r}}/B_{\mathrm{r}}}{\Delta T} = ( -330 \pm 20 ) \times 10^{-6} \,\mbox{K}^{-1}
\end{equation}
at a temperature of $21.5\,^\circ$C.

\section{Measurement of the vertical gradient of the radial field}

One of the key objectives in designing this magnet was to obtain a flat field profile, i.e., a small change of the radial field as a function of the vertical position. In other words, the vertical gradient of the radial field should be as small as possible. There are two reasons for this objective: First, the force mode consists of two different measurements called mass-on and mass-off. Between the two measurements, the coil position changes slightly in vertical position. If the field profile is flat, the flux integral remains the same  for both measurements and no correction is required. Second, during the velocity mode the coil is moved through the magnetic field such that the induced voltage stays constant. In a flat field profile, the velocity required to achieve constant induced voltage remains constant and is thus easier to measure. A flat field profile reduces uncertainties in watt balance experiments. The goal was that the magnetic flux density should vary by less than $\pm 0.01\,\%$ over the inner 8\,cm of the gap.

Two methods were employed to measure the vertical gradient of the radial field: a guided Hall probe and a gradiometer coil. Two setups were used for the Hall-probe method, one built by EEC and the other by NIST.  In both systems, a brass tube is centered on the gap of the magnet in which a second brass tube containing a Hall probe (Lakeshore MMZ-2518-UH and HMMT-6704-VR for the EEC and NIST system, respectively) is guided. The guide tube was centered in the air gap by two tapered Teflon plugs, one at the top and one at the bottom. The probe was centered to be concentric with one access hole each at the upper and lower yoke cap. The guide tube was mounted in a coaxial hole in both Teflon cones. The magnetic field was recorded at different positions. The EEC setup required manual vertical positioning of the Hall probe, while the NIST setup used a motorized translation stage. The resolution of each Hall probe is $1\times10^{-4}$\,T, which corresponds to a relative change in the magnetic flux density of $2\times 10^{-4}$. In order to measure smaller changes in the field, multiple measurements have to be averaged.  We averaged the field profile measured in all twelve holes to one profile. This procedure discards the azimuthal information and obtains an average vertical profile.

The gradiometer coil consists of two identical coils wound on a single former displaced in the vertical direction. Each coil has $N=464$ turns and a mean radius of $r=217.5$\,mm. The height of each coil is 10\,mm and the centers of the coils are displaced by $\Delta z=11.5$\,mm. The two coils are electrically connected in series opposition. Two voltmeters are used to measure the induced voltages as the coil assembly is moved with constant velocity, $v\approx 2$\,mm/s through the magnet. One voltmeter measures the voltage induced in one coil, the other the difference. The ratio of the two measurements is given by
\begin{equation}
\frac{V_1(z)-V_2(z)}{V_1(z)} = \frac{B_{\mathrm{r}}(z)-B_{\mathrm{r}}(z-\Delta z)}{B_{\mathrm{r}}(z)} \approx \frac{\Delta z \frac{\displaystyle \mbox{d}B_{\mathrm{r}}(z)}{\displaystyle \mbox{d}z}}{B_{\mathrm{r}}(z)}. \label{eq2}
\end{equation}
The absolute magnitude of the radial magnetic flux density can be estimated from the mean velocity, $v\approx 2$\,mm/s of the coil and the coil's dimensions using  $\bar{B_{\mathrm{r}}}=\bar{V}_1/(v N 2\pi r)$. Vibrations induced by the coil motion cause excess noise on $V_1$ with several mV amplitude. To get a good estimate of the magnetic flux density, the voltage was averaged over the central 80\,mm. Note that the mean value is not the important quantity in this measurement.

\begin{figure}[h!]
\centering
\includegraphics[width=3.5in]{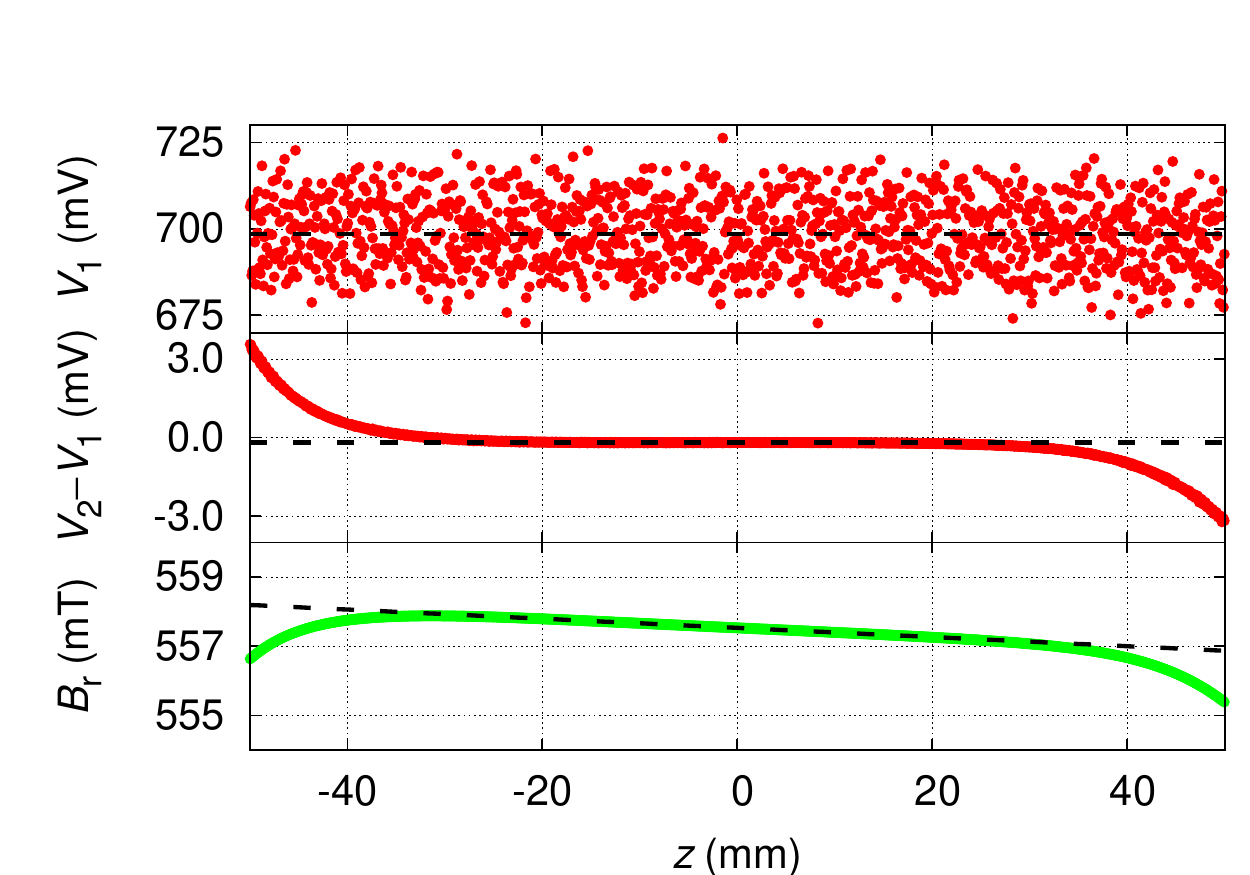}
\caption{Measurements with the gradiometer coil. The top two graphs show the raw data, $V_1$ and $V_2-V_1$. The bottom graph shows the radial flux density as a function of position calculated from the raw data. The horizontal axis is such that zero is the center of the magnet and positive numbers are above the center.}
\label{fig:onesweep}
\end{figure}

The vertical variation of the field is calculated by numerically integrating equation~\ref{eq2} yielding
\begin{equation}
B_{\mathrm{r}}(z) = \frac{ \bar{B_{\mathrm{r}}}} {\bar{V}_1 \Delta z } \int_b^z \left( V_1(z\prime)-V_2(z\prime)\right)\mbox{d}z\prime + O,  \label{eq3}
\end{equation}
where $O$ is chosen such that $B_{\mathrm{r}}(0) = \bar{B_{\mathrm{r}}}$. Since both coils are mounted on the same coil former and are immersed in approximately the same flux density, the voltage noise on the difference is reduced by a large factor (about 1000). Fig.~\ref{fig:onesweep} shows a typical measurement. In the central region, the difference of $V_1$ and $V_2$ is $-0.19$\,mV, indicated by the dashed line in the middle plot of the figure. This voltage difference corresponds to a slope in the field of -13\,\mymu T/mm. To exclude systematic errors, i.e., caused for example by a coil winding error, we performed one measurement with the coil mounted up-side down. After correcting for the electrical connections, we obtained the same field profile.

The gradiometer coil was preferred over the guided Hall probe to measure the field profile with high accuracy because of several reasons. The measurement with the gradiometer coil is first order independent of the concentricity of coil and magnet. The result is also in first order insensitive to the parallelism of the motion axis to the magnet axis. Furthermore, the coil integrates the field along the azimuthal direction. The gradiometer coil measurement has enough resolution to measure even small field gradients.  During the construction of the magnet, measurements with the gradiometer coil were performed twice. The manufacturer used the guided hall probe to measure the field profile. As it is detailed below, the attempts to shim the field by grinding the outer yoke did not converge. This was not due to limitations of the field measurements. It was, as we learned later, due to the change of the field profile caused by opening and closing the magnet.

\section{Initial assembly and attempts to shim the field}
\begin{figure}[h!]
\centering
\includegraphics[width=3.5in]{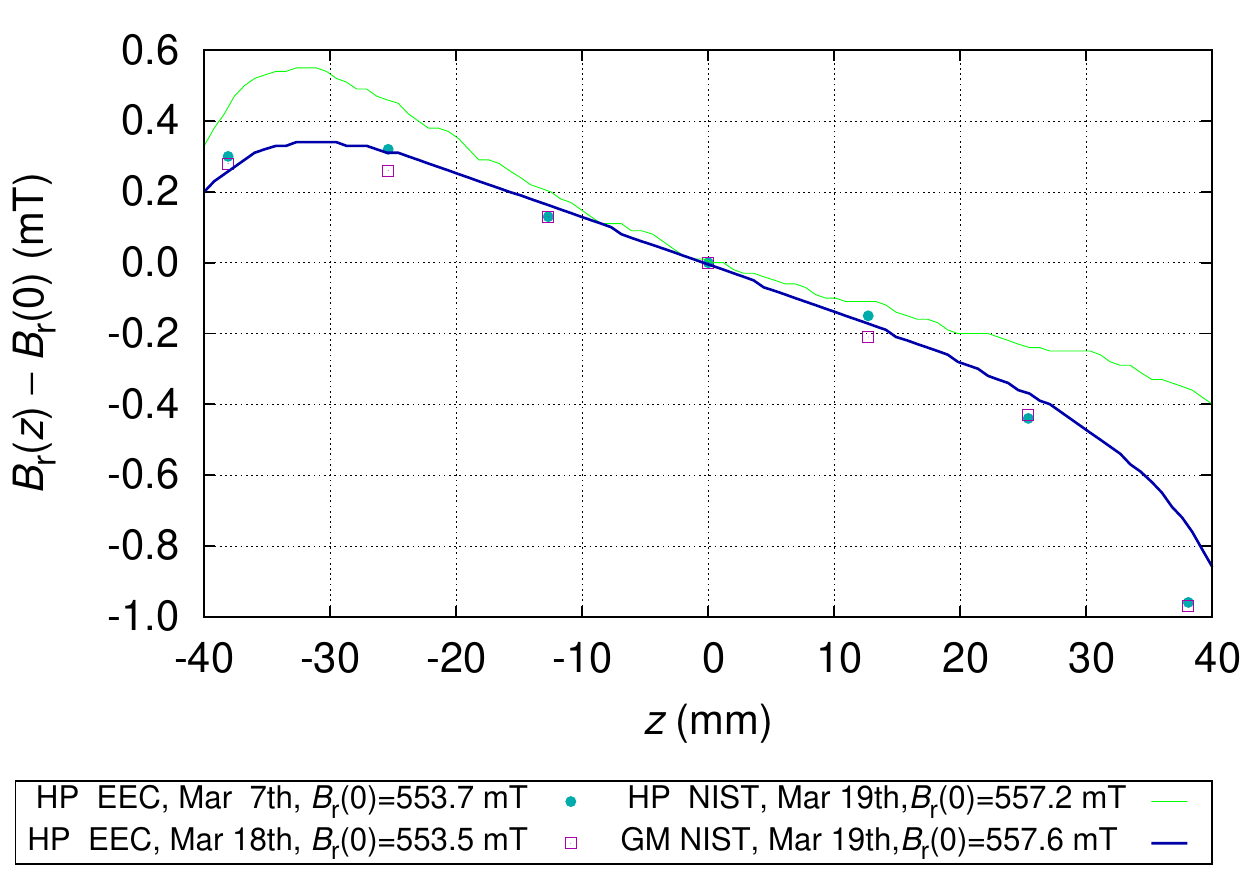}
\caption{Four measurements of the radial flux density as a function of vertical position after the magnet has been assembled for the first time. The measurements performed with the Hall probe are abbreviated by HP, the gradiometer coil by GM.}
\label{fig:compare}
\end{figure}

After all pieces of the magnet were manufactured, the magnet was assembled for the first time and the radial magnetic flux density of the magnet was measured. This measurement was performed at the manufacturer's facility with three different methods. Besides the gradiometer coil, two Hall probes were used.  The measurement with the EEC Hall probe was performed on two different days about 1.5 weeks apart.  The results of the measurements are shown in Fig.~\ref{fig:compare}. In order to overlay the measurements, the value of $B_{\mathrm{r}}(0)$ has been subtracted from each measurement.  All four measurements show a similar slope of the radial magnetic flux density, about -13\,\mymu T/mm, which is about a factor of 10 larger than intended. 

The  measured variation of the radial flux density in the precision gap of at least 1\,mT failed the requirement of $\Delta B_{\mathrm{r}}/B_{\mathrm{r}}< 2 \times 10^{-4}$ by a factor of 10. Based on this measurement, it was decided to regrind the inner diameter of the outer yoke (part 2 in Fig.~\ref{fig:circuitClose}). The specification for this regrinding was to add a taper such that the gap is nominally 3.000\,cm at the top to 3.008\,cm at the bottom. Varying the gap is a known technique to engineer a desired field profile~\cite{bnm,eichenberger04}.

\begin{figure}[h!]
\centering
\includegraphics[width=3.5in]{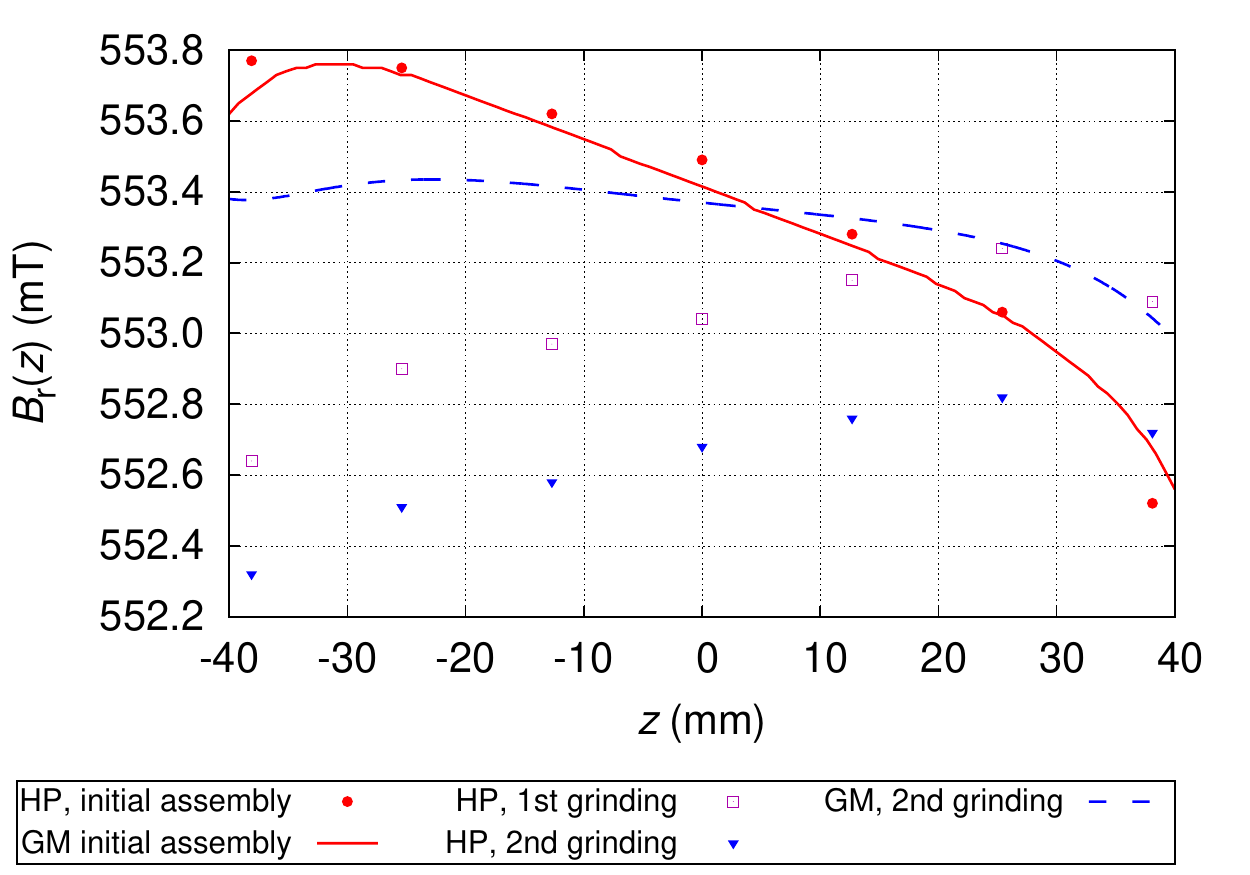}
\caption{The points represent data measured with the EEC Hall probe (HP) after initial assembly, the first, and second grinding process. The lines indicates measurements performed with the gradiometer coil (GM) after initial assembly and after the second grinding. The gradiometer coil readings are 4.2\,mT higher. For this plot 4.2\,mT, were subtracted from the measurements performed with the gradiometer coil. }
\label{fig:grindingresult}
\end{figure}

Fig.~\ref{fig:grindingresult} shows the measurement of the radial magnetic flux density after grinding the outer yoke. This measurement was performed only at EEC with their Hall probe. The slope of the radial magnetic flux density at the center has changed from -13\,\mymu T/mm to 7\,\mymu T/mm. From this measurement, it was concluded that the grinding overshot by approximately 50\,\%. The outer yoke was sent back to the grinding house with the instruction to grind the gap such that it is  nominally 3.003\,cm at the top to 3.008\,cm at the bottom, reversing 1/3 of the first grinding process. After the second grinding process, the magnetic flux density was measured again at EEC. This result was almost identical to the previous measurement. From this, it was concluded that the measurements with the Hall probe are not reliable at this level. It is possible that the trajectory of the Hall probe was not centered well enough on the gap. For example, to measure a slope in the radial magnetic flux density of 7\,\mymu T/mm in a perfectly uniform field, the probe only needs to travel sideways by 0.24\,mm over the 8\,cm region. While the probe was certainly positioned better than 1\,mm in the center of the gap, an accuracy of 0.2\,mm could not be ensured. After the second grinding, the gradiometer coil was brought to EEC to remeasure the profile of the radial magnetic flux density. A slope of -3.5\,\mymu T/mm was observed, see Fig.~\ref{fig:grindingresult}.

Since the grinding process did not seem to converge to a flat field profile, other shimming techniques were explored. The first approach was to insert low carbon steel rods in the inner diameter of the lower \smco\ ring. As can be seen from Fig.~\ref{fig:grindingresult}, the flux density was larger at the lower part of the magnet (negative $z$ values). Inserting iron in the ring changed the slope of the radial magnetic flux in the center of the magnet by approximately 1\,\mymu T/mm, which was a factor of three smaller than needed. Hence, this strategy was abandoned. 

\begin{figure}[h!]
\centering
\includegraphics[width=3.5in]{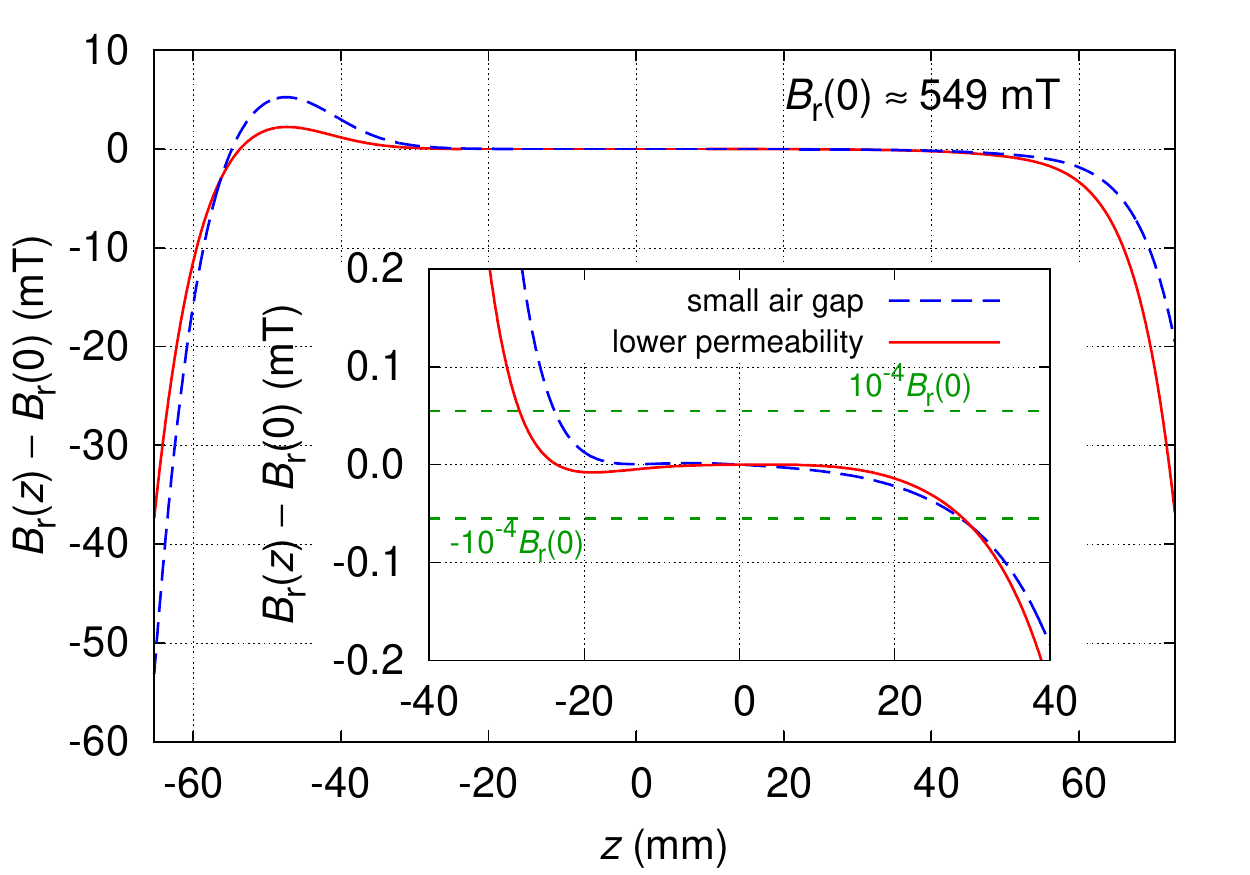}
\caption{The radial magnetic flux density as a function of vertical position. The inset shows the central region magnified such that the total size of the vertical axis extends $\pm 4\times 10^{-4}$ around the value at $z=0$. Two shimming methods lead to a similar profile: introducing a small air gap (dashed line) and reducing the relative permeability of the iron (solid lines).}
\label{fig:good}
\end{figure}

A better way to shim the field is to introduce a small air gap between the lower third of the magnet and the upper two thirds, i.e., a gap between the pieces 2 and 6 on the top and the pieces 3 and 7 on the bottom in Fig.~\ref{fig:circuitClose}. A flat profile is obtained when this additional air gap is about 0.5\,mm high. A stable and uniform air gap can be achieved by inserting aluminum shim stock pieces at several azimuthal locations. This small air gap increases the reluctance of the lower part of the yoke. Hence, the lower \smco\ ring contributes less flux to the  magnetic flux density of the gap. The profile that is obtained with this method is shown as the dashed line in Fig.~\ref{fig:good}. While this shimming method obtains a flat profile, it has one disadvantage: A small air gap connects the precision air gap inside the magnet to the outside world and flux leaks out of the magnet. Hence, the shielding of the magnet is compromised. 

\begin{figure}[h!]
\centering
\includegraphics[width=3.5in]{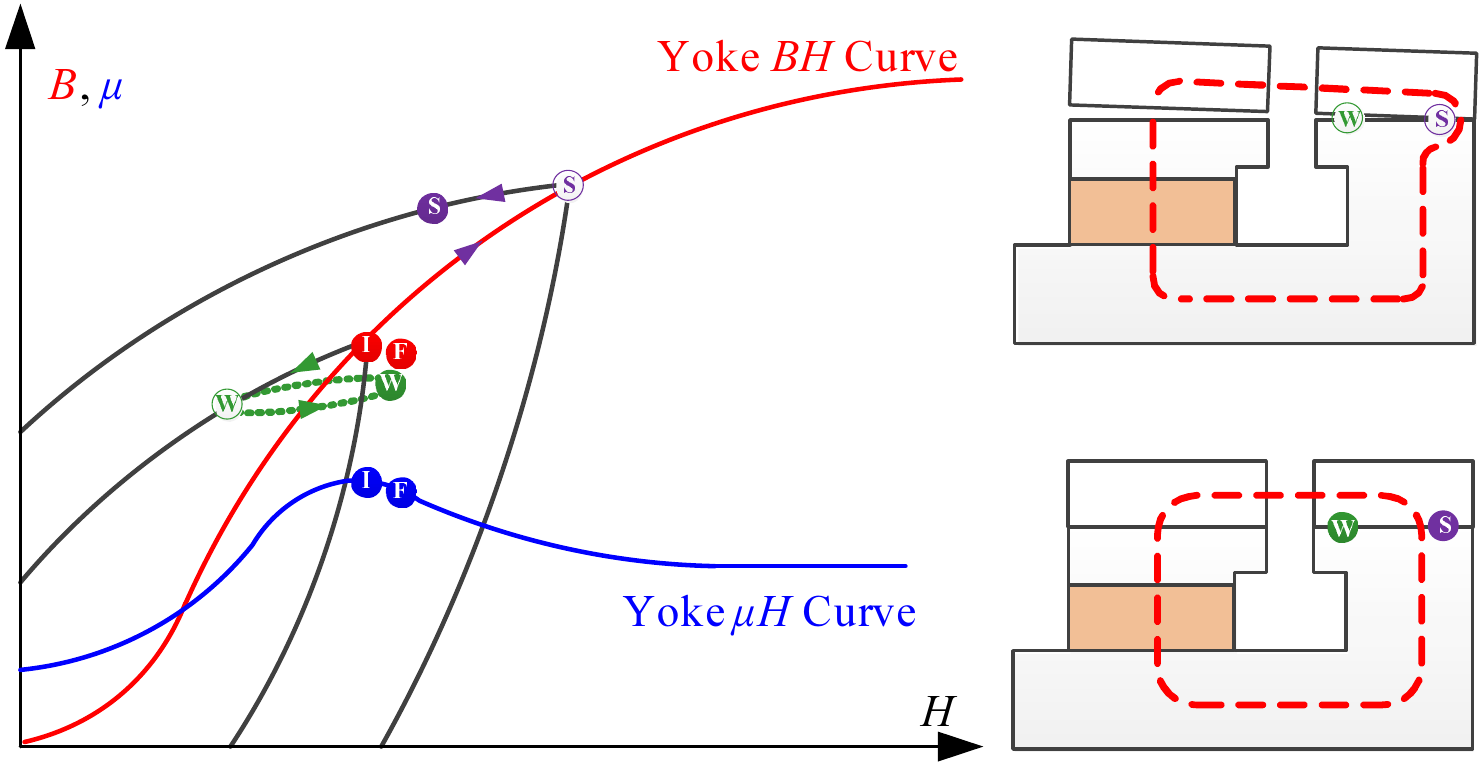}
\caption{Schematic drawing showing the change in the outer part of the yoke before, during, and after opening with an angle. We consider the locations S and W at the small and wide side of the gap on the outer yoke. At the beginning, the magnetic state is given by the point I on the $B$-$H$ curve. During opening, the  point W moves along a minor hysteresis loop to smaller values of $B$ and $H$. The point S moves along the major curve to a higher value of $B$ and $H$. After closing, the point W moves almost back to the original point (the solid W). The point on the small side of the wedge moves along a minor loop to a new point (solid S) that has substantial more $B$. The mean value can be found at the point denoted by F for final point. Overall the $\mu_{\mathrm{r}}$ of the yoke has decreased, reducing the flux in the air gap.
}
\label{fig:hyst_exp}
\end{figure}

We noted that the slope in the center of the gap changed by a few \mymu T/mm every time the magnet was opened and closed. An examination of this effect yielded another shimming strategy. The variability in the vertical linear gradient of the magnetic flux density is caused by non parallel opening and closing of the magnet. In this case, a situation occurs where the lower part of the yoke touches the upper part of the yoke on one spot along the outer circumference. A large amount of flux is driven through this contact point, see Fig.~\ref{fig:hyst_exp}. This effectively shifts the working point of the iron at the contact zone on the $B$-$H$ curve to the right, i.e, to a point with smaller relative permeability. Even after the magnet is closed, the iron remains in a state of smaller relative permeability due to the hysteretic behavior of the $B$-$H$ curve. Hence, in the closed state this part of the yoke conducts the magnetic field less well and the flux in the gap is lower.

The shimming process works as follows:  (1) The magnet is opened by a little more than 1\,mm. (2) A 0.5\, mm thick shim piece with a size of approximately 5\,cm by 5\,cm is inserted in the 1\,mm gap at an azimuthal position $\alpha$. (3) The magnet is closed. Due to the shim, the magnet closes in a tilted fashion and the iron at the azimuthal position $\alpha+180^\circ$ is driven to the state with less relative permeability. Steps (1) through (3) are performed a total of six times, where the azimuthal position is advanced by $60^\circ$ every time. After this, the iron is at the less permeable state for the entire circumference.

This shimming process is repeatable. We were able to reproduce the shimming procedure several times, yielding an almost identical field profile.

We have two concerns using this shimming procedure: How stable is the field in the gap obtained with this method? Does this process change the azimuthal symmetry of the field? We measured the field profile over 3 days every 30 minutes and we found that the slope of the radial magnetic flux density changed linearly with time from -0.594~\mymu T/mm to -0.609~\mymu T/mm over 60 hours. Hence the slope changes with a rate of  $2.5\times10^{-10}$\,T/(mm\,h). This is enough stability for a watt balance experiment, where the  flux integral is measured every hour. The azimuthal variation of the magnetic flux density is hard to measure with high precision. It can only be measured with the Hall probe, since coils integrate over the azimuthal dependence. Using the Hall probe, however, requires precise positioning along the radial direction inside the gap. To compare the magnetic flux density at two azimuthal angles, the Hall probe must be positioned at the center of the gap through different access holes in the top of the magnet. A difference in probe placement of 1\,mm causes a different measurement of 2.3\,mT. The measurements before and after the shimming performed through all twelve access holes  ($30^\circ$ increments) showed a similar maximum difference of 1.5\,mT. This difference could be due to a real field inhomogeneity or due to a positioning error. Within the measurement uncertainty, the shimming procedure did not make the azimuthal asymmetry worse.

In summary, a flat profile of the magnetic flux density as a function of vertical position can be achieved with two different shimming methods. One can introduce a small air gap between the lower and upper part of the magnet or lower the permeability of the iron yoke in the lower half of the magnet by exposing it to a large magnetic field. Both methods increase the reluctance of the flux path around the lower \smco\ ring. Fig.~\ref{fig:good} shows the field profile achieved with both methods. With these two methods, a slope of less than 0.1\,\mymu T/mm or in relative terms $2\times 10^{-7}$/mm, can be achieved. The relative flux density stays between $\pm 1\times 10^{-4}$ over at least 5\,cm. This is a bit less than the initial goal of 8\,cm. We plan on using the shimming method that decreases the permeability of the yoke for the first watt balance measurements with this magnet.

\section{The radial dependence of the magnetic flux density}

The insight that a $1/r$ dependence of the radial flux density allows the construction of a better watt balance is attributed to P.T.~Olsen. His argument goes as follows: Assume that 
\begin{equation}
B_{\mathrm{r}}(r)=B_{\mathrm{o}} \frac{r_o}{r},
\label{eq:radial}
\end{equation}
the coil is centered on the magnetic field, has a radius $r$, and $N$ turns. The flux integral is given as a line integral along the wire,
\begin{equation}
f(r) \equiv Bl =\hat{z} \int_{\partial S} \vec{B} \times \mbox{d}\vec{l}.
\end{equation}
Assuming no azimuthal dependence of the field, this integral equates to $f(r) = 2\pi N B_{\mathrm{r}}(r) r$. 
For the field given in (\ref{eq:radial}), the flux integral evaluates to $f=2\pi N B_{\mathrm{o}} r_{\mathrm{o}}$, which is independent of $r$. In other words, for a $1/r$-field, $\mbox{d}f/\mbox{d}r=0$. 

One important assumption in the watt balance experiment is that the flux integral in the force mode is identical to the one in the velocity mode. In the force mode current is passed through the coil which leads to heating and subsequently thermal expansion of the coil. If the flux integral is independent of the coil radius $r$, the above assumption holds. If this is not the case, a bias is introduced into the experiment.

In order to investigate the deviation from a perfect $1/r$-field, it is useful to expand the flux integral for small changes in radius and assume $\mbox{d}f/\mbox{d}r\neq0$. In this case,
\begin{equation}
f(r+r\gamma)  \approx f(r) +r\gamma \frac{\mbox{d}f}{\mbox{d}r},  \label{eq:f-taylor}
\end{equation}
where $\gamma\ll 1$. The last term can be rewritten as 
\begin{equation}
r\gamma \frac{\mbox{d}f}{\mbox{d}r} = \gamma \beta f \;\;\;\mbox{with}\;\;\;\beta = \frac{r}{f}\frac{\mbox{d}f}{\mbox{d}r}. \label{eq:beta}
\end{equation}
Here, $\beta$ is a unitless number describing the deviation of the flux density from a $1/r$ dependence. 

To measure the radial dependence of the radial flux density, a radial gradiometer coil was built. This gradiometer coil consists of three coils on a single former. Each coil has 295 turns, a vertical size of 17\,mm and a radial width of 4.9\,mm. The mean radii of the three coils are $r_{\mathrm{i}}=211.45$\,mm, $r_{\mathrm{m}}=216.35$\,mm, and $r_{\mathrm{o}}=221.25$\,mm. For the measurement, the inner and outer coil  are connected in anti-series to one voltmeter and the middle coil to a second voltmeter. Both voltmeters are sampled at the same time, while the gradiometer coil is vertically moving through the gap of the magnet with a velocity of $v=2$\,mm/s. A value of $\beta$ is estimated using

\begin{equation}
\beta \approx \frac{V_{\mathrm{o}-\mathrm{i}}}{V_{\mathrm{m}}} \frac{r_{\mathrm{m}}}{r_{\mathrm{o}}-r_{\mathrm{i}}}.
\end{equation}

\begin{figure}[h!]
\centering
\includegraphics[width=3.5in]{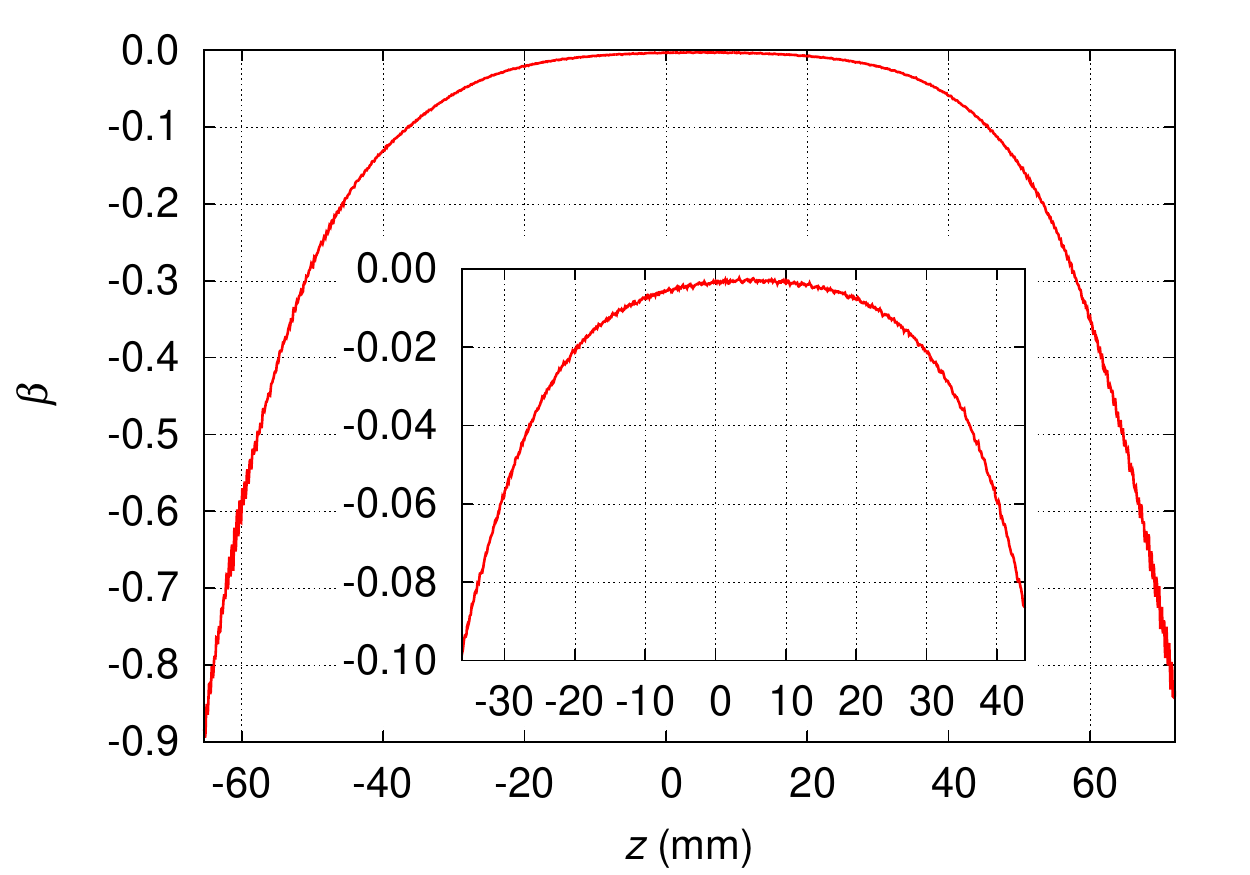}
\caption{Measurement of the deviation of the radial field from $1/r$, see text for a definition of the unitless constant $\beta$. The measurement was performed at $r_{\mathrm{m}}=0.216$\,m.}
\label{fig:radial}
\end{figure}

Fig.~\ref{fig:radial} shows the measurement for $\beta$ around the symmetry plane of the magnet, $\beta=-0.003$. The negative sign indicates that the field drops faster than $1/r$ with increasing radius. The same value of $\beta$ is obtained when the middle and outer coil or the inner and middle coil are combined. Since we cannot completely rule out other sources of induced electromotive force, we conservatively interpret the measured value as an upper limit for $\beta$, i.e., $|\beta|<0.003$. Using this $\beta$, the change in flux integral  due to a geometry change caused by e.g., coil heating, can be calculated. Changing the temperature of the coil by $\Delta T$ causes a radial expansion of $\Delta r = r \alpha \Delta T$, where $\alpha=16.6\times 10^{-6}$\,K$^{-1}$ is the linear coefficient of expansion for copper. Note, that $\gamma = \alpha \Delta T$. According to (\ref{eq:f-taylor}) and (\ref{eq:beta}) the relative change in flux integral is $\gamma\beta=5\times 10^{-8}$\,K$^{-1} \Delta T$. In the current design of NIST-4 the power dissipation in the force mode is about 8\,mW (R=130\,$\Omega, I=8\,$mA). Assuming a copper mass of 3\,kg the temperature of the coil would rise by 0.026\,K in one hour. Note this estimation neglects losses in thermal energy due to radiation to the environment. The corresponding relative change in the flux integral would be $1.3\times 10^{-9}$.  To  further minimize this effect, a coil heater can be installed as was done in the NPL watt balance~\cite{Robinson12}. The deviation from a $1/r$ field gets rapidly worse with increasing distance of the coil to the symmetry plane of the magnet. At a distance of 2.2\,cm, $\beta$ is already 10 times larger.

\section{Measurement of the reluctance force}
As discussed in~\cite{ss13}, the reluctance force pulls the coil into the center of an iron structure like the yoke of this magnet regardless of the sign of the current in the coil. The force originates from the fact that a current carrying coil has minimum energy in the center of the yoke. Note, this effect is independent of the magnetic field. If the \smco\ rings were replaced by magnetically inactive stainless steel rings ($\mu_{\mathrm{r}}=1$), the effect would still be present. The reluctance effect is similar to the effect exploited by a solenoid actuator, where an iron slug is pulled into a solenoid after it has been energized. 

The energy of the magnetic field produced by the coil is given by $E=(1/2)LI^2$. From the energy, the vertical force can be calculated using
\begin{equation}
F_{\mathrm{z}} = \frac{\mbox{d}E}{\mbox{d}z}=\frac{1}{2} I^2 \frac{\mbox{d}L}{\mbox{d}z}
\end{equation}
assuming that the current in the coil is maintained at a constant level. In order to estimate this effect, the inductance of the coil, $L$ has to be measured as a function of vertical position in the magnet, $z$.

The measurements below were carried out by connecting a precision $50\,\Omega$ resistor in series with the vertical gradiometer coil. This time, the two coils on the gradiometer coil were connected in series to form effectively one coil with 928 turns. A sinusoidal voltage with an amplitude of 2\,V and frequency, $f$, was applied. Two Agilent 3458A voltmeters were used to  simultaneously measure the voltage across the 50\,$\Omega$ resistor and the coil. Fitting sines to both of these measurements yielded the amplitudes and the relative phase between these two measurements. From the amplitudes and the phase  difference, the electrical resistance and the inductance of the coil could be reconstructed.

The inductance of the coil is  measured at $f$, yielding $L(f)$. Since the watt balance operates near DC, we are interested in $\lim\limits_{f \to 0} L(f)$. We placed our gradiometer coil in the center of the magnet and measured $L(f)$ using the procedure explained above. We found that for low frequencies ($f<1$\,Hz) the inductance scales like $L(f)=a-b\sqrt{f}$, with $a=4.2$\,H and $b=0.66$\,H/$\sqrt{\mbox{Hz}}$. The same coil outside the magnet and far away from any metal has an inductance $L=0.8$\,H , which is independent of $f$ for $f<100$\,Hz. The frequency dependence of the inductance of the coil inside the magnet is due to the skin effect~\cite{Gonzales65}. In solid iron, the skin depth is very small since it has a high conductivity and a high susceptibility. At 1\,Hz, the skin depth is about 5\,mm. Hence in order for the field to completely penetrate the inner yoke, $f$ has to be below 0.6\,mHz.

\begin{figure}[h!]
\centering
\includegraphics[width=3.0in]{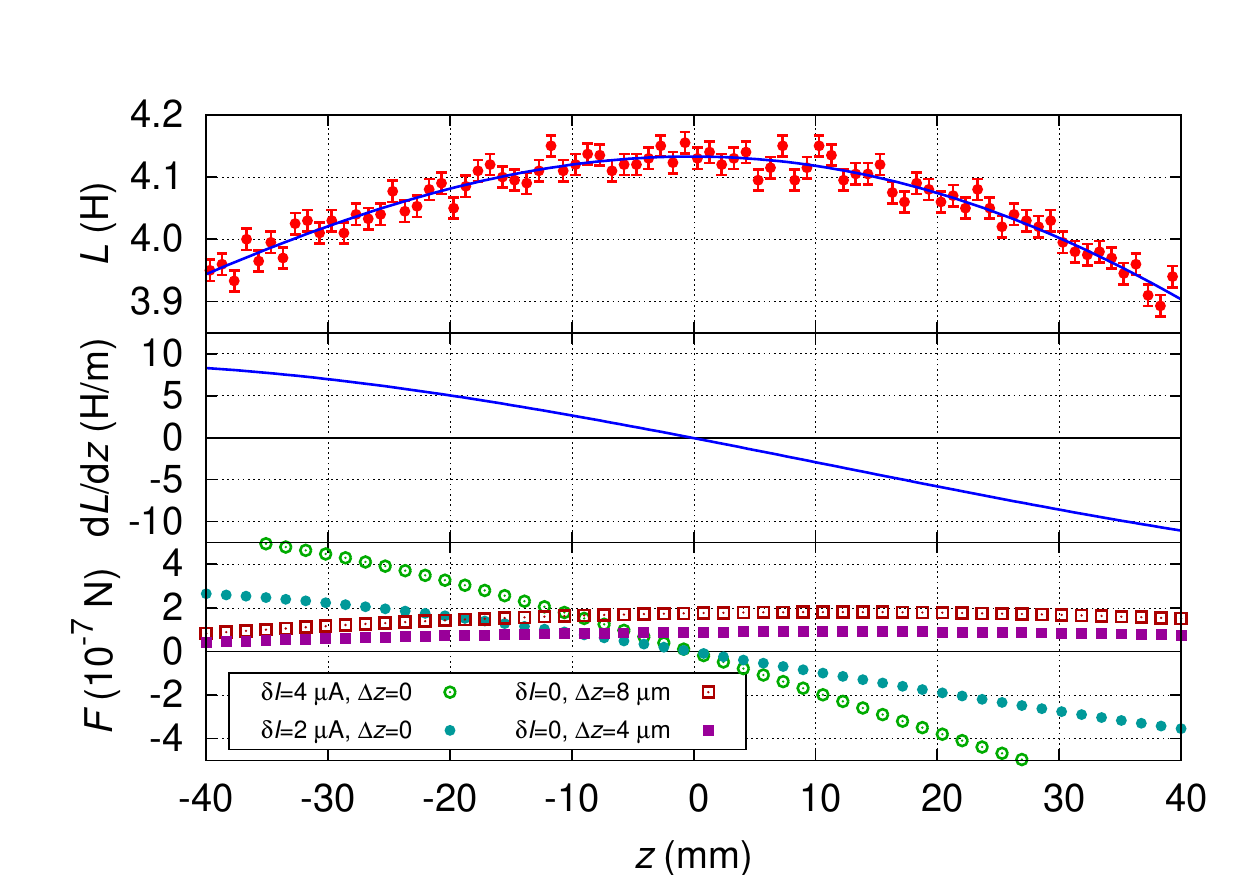}
\caption{The top graph shows the measurement of the inductance, $L$, of the gradiometer coil (928 turns).  $L(z)$ is measured at $10$\,mHz.  Each point represents an average value of two measurements. The error bars are obtained from the fit. The solid line is a fit to the data using a fourth order polynomial. The middle graph shows the derivative calculated from the polynomial fit. The bottom graphs show the results of (\ref{eq:spur}) for four different cases. For these calculations $\bar{I}=8$\,mA is assumed.
}
\label{fig:L_of_z}
\end{figure}

To obtain a good estimate for $\lim\limits_{f \to 0} L(f)$, the measurement was carried out at $f=0.01$\,Hz. In this case, the deviation from the DC value is at most $0.06$\,H and we found no position dependence of this difference. $L(z)$  was measured for every mm along the $z$ direction. The result is shown in the top graph of Fig.~\ref{fig:L_of_z}. In the middle graph of the figure, $\mbox{d}L/\mbox{d}z$ is plotted. This derivative is calculated from a fourth order polynomial fit to the raw data. The second derivative of the inductance is mostly independent of $z$ and evaluates to $\mbox{d}^2L/\mbox{d}z^2 = -346$\,H/m$^2$ at the center of the magnet.

The spurious force signal to the watt balance experiment due to the reluctance effect can be written as
\begin{equation}
F = \frac{1}{2} I_{\mathrm{on}}^2 \left.\frac{\mbox{d}L}{\mbox{d}z}\right|_{z=z_{\mathrm{on}}} - \frac{1}{2} I_{\mathrm{off}}^2 \left.\frac{\mbox{d}L}{\mbox{d}z}\right|_{z=z_{\mathrm{off}}},
\end{equation}
where $I_{\mathrm{off}}$, $I_{\mathrm{on}}$ and $z_{\mathrm{off}}$, $z_{\mathrm{on}}$ are the currents and positions  of the coil  during the mass-off and mass-on measurement, respectively.

This equation simplifies in first order to 
\begin{equation}
F \approx  2 \delta I I_\mathrm{A}  \frac{\mbox{d}L(\bar{z})}{\mbox{d}z} - (I_\mathrm{A})^2 \Delta z \frac{\mbox{d}^2 L(\bar{z})}{\mbox{d}z^2}, \label{eq:spur}
\end{equation}
where  $\bar{z}=(1/2) (z_{\mathrm{on}}+z_{\mathrm{off}})$, $\Delta z=(1/2) (z_{\mathrm{on}}-z_{\mathrm{off}})$, $\delta I=(1/2) (I_{\mathrm{on}}+I_{\mathrm{off}})$, and $I_\mathrm{A} = (1/2) (I_{\mathrm{on}}-I_{\mathrm{off}})$. Typically a watt balance is operated such that  $ I_{\mathrm{on}}\approx -I_{\mathrm{off}}$, hence $\delta I\approx0$ and $I_\mathrm{A} \approx I_{\mathrm{on}}$. The results of the above equation for four different cases are shown in the lower plot of Fig.~\ref{fig:L_of_z}.

%
%
%

The spurious force needs to be compared to the force that will be generated by the watt balance, i.e. $\approx 10$\,N. In order to keep the relative contribution of the spurious force to the measurement below $10^{-8}$, a maximum force of $F_{\mathrm{m}}=10^{-7}$\,N is permitted. We will use $10^{-7}$\,N as a benchmark for the analysis below

The first term on the right of (\ref{eq:spur}) can be made small, even for a finite $\delta I$, by performing the watt balance experiment at the center of the yoke, where $\mbox{d}L/\mbox{d}z=0$. The slope at which the spurious force changes with deviations from the ideal position depends on the current mismatch, $\delta I$, as is shown in Fig.~\ref{fig:L_of_z}. For a current mismatch of $\delta I=4\,\mymu$A  the range of operation where $|F|<F_{\mathrm{m}}$is $\pm4.5$\,mm.

The second term remains approximately constant for different coil positions, since the second derivative of the inductance with respect to $z$ is largely independent of $z$. It evaluates to $\Delta z \times 2.2\times 10^{-8}\,$N/\mymu m at the center of the magnet. In order to keep the absolute value of this term smaller than $F_{\mathrm{m}}$, the change in coil position, $z_{\mathrm{off}}-z_{\mathrm{on}}$ must be smaller than 9\,\mymu m. Typically, in an experiment like NIST-4, the coil position between the mass-on and mass-off state can be maintained within a few micrometers of each other. In this case, the second term of (\ref{eq:spur}) is about $F_{\mathrm{m}}/3$. This number is certainly large enough to be considered as systematic uncertainty of the experiment. However, it will not be a dominant effect.

We would like to emphasize that it is important to measure the inductance of the coil at low frequency. We performed the same measurement with $f=100$\,Hz. Using this data, one would calculate a reluctance force that is about 10 times smaller than the real reluctance force at DC. 

Equation~10 in~\cite{ss13} gives an order of magnitude estimation of the reluctance force. Using this estimation to calculate  $\mbox{d}^2L/\mbox{d}z^2$, a value of -800\,H/m$^2$ is found. While the absolute value of this number is almost twice as large as the result obtained from the measurement, the order of magnitude is right, as intended.

\section{The magnetic field outside the magnet}
One concern of any magnet system being developed for a watt balance is the magnetic field at the location of the mass. The interaction between the field and the mass can create a spurious force that may lead to a systematic effect. In general, the vertical force on an object with a volume susceptibility $\chi$ and permanent magnetization $\bf{M}$ is given by
\begin{equation}
F_{\mathrm{z}}=-\frac{\muz}{2}\frac{\partial}{\partial z}\int \chi \bf{H}\cdot\bf{H}\mbox{dV}-\muz \frac{\partial}{\partial z}\int \bf{M} \cdot\bf{H}\mbox{dV},
\label{eq:mag}
\end{equation}
see, e.g.,~\cite{davis95}. 

\begin{figure}[h!]
\centering
\includegraphics[width=3.5in]{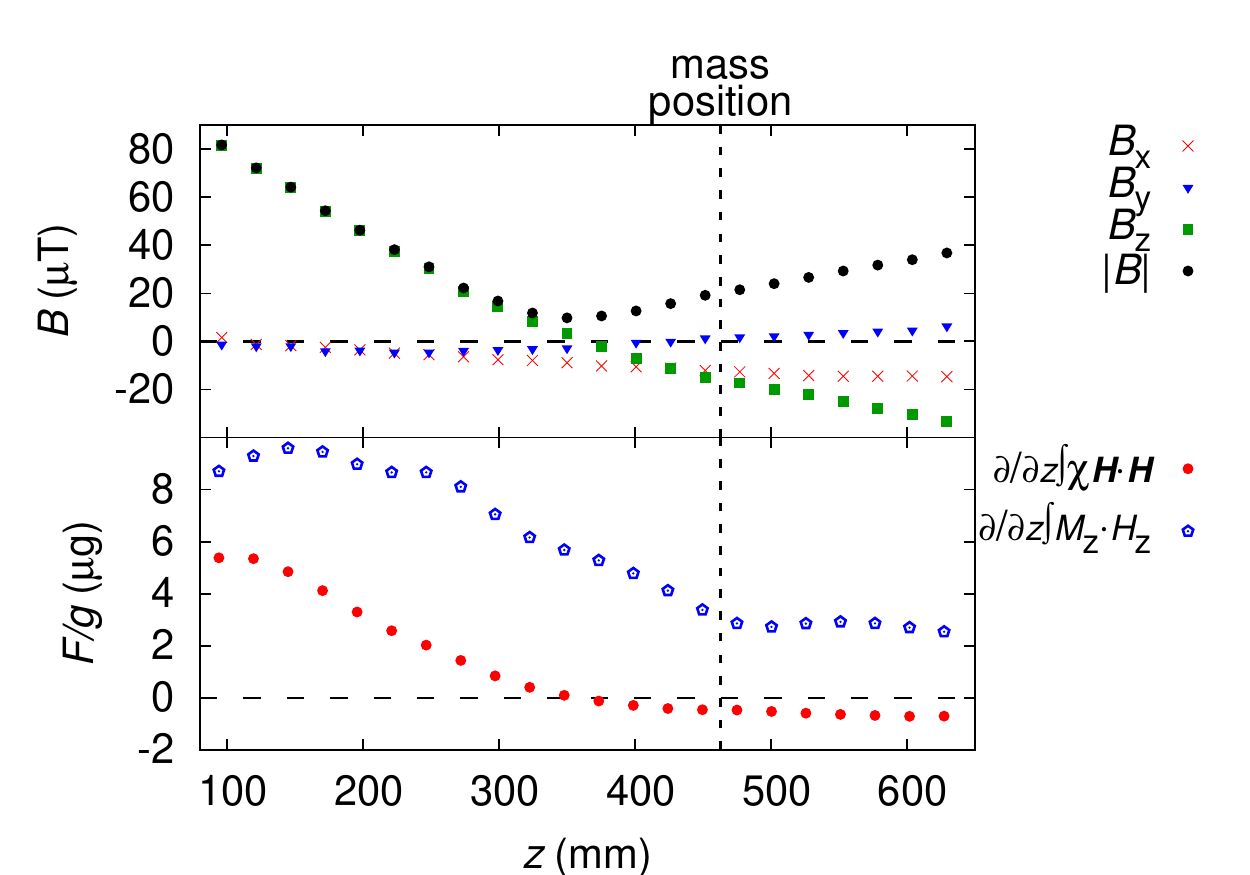}
\caption{The top plot shows the three components and the absolute magnitude of the magnetic flux density above the magnet as a function of the distance above the top surface. The bottom graph shows the systematic force divided by the local acceleration caused by the magnetic susceptibility and the magnetization of the weight.  For this calculation, a 1\,kg E$_1$ weight was assumed with the maximum allowed values for the magnetic susceptibility, $\chi=0.02$ and magnetization, 2\,A/m. The vertical dotted line indicates the position of the mass during weighing.}
\label{fig:field_force}
\end{figure}

The three components of the flux density above the permanent magnet system have been measured as a function of distance from the top surface. This measurement was performed near the symmetry axis using a 3 dimensional magneto resistive sensor. In Fig.~\ref{fig:field_force},  the three components and the absolute magnitude are shown. At close distances, the vertical component of the field is dominant. It decreases in a nearly linear fashion with growing distance until it vanished at a distance of 350\,mm. From there on, it decreases further to match the vertical component of the ambient field, about 45\,\mymu T. The horizontal components are close to zero in the first 300\,mm. At larger distances, they approach the ambient values.

In order to calculate the force from these measurements, few simplifications were made: a 1\,kg stainless steel weight with a height of $69.1$\,mm and a diameter of $48$\,mm was assumed. The magnetic susceptibility was assumed to be constant over the volume of the weight and independent of $H$, which is reasonable at these small fields. The magnetization is also assumed to be constant over the volume of the weight. For simplicity, we assumed two components of the magnetization to be zero, i.e., $M_{\mathrm{x}}=M_{\mathrm{y}}=0$. Ideally, NIST-4 is able to realize mass using an E$_1$ class weight. In order to calculate the worst case force on such a weight, we assume the maximum permissible limits for the susceptibility and the magnetization. According to ~\cite{oiml111},  $\chi=0.02$ and $M_{\mathrm{z}}=2$\,A/m were assumed. 

The bottom plot in Fig.~\ref{fig:field_force} shows the calculated magnetic force for the two terms in (\ref{eq:mag}). The first term gives the force due to the magnetic susceptibility of the mass. It depends on the derivative of the squared magnitude of the magnetic field. Since the magnetic field has a minimum around 350\,mm, this term changes sign at this point. The second term gives the force due to a permanent magnetization of the mass. This force depends on the derivative of the $z$-component of the magnetic field. The derivative is negative for the entire data stretch leaving a positive force on the mass. The magnitude of the force generated by the first term is smaller than 1\,\mymu g for $z>$300\,mm and can be neglected. The second term produces a force that can be as large as 3\,\mymu g for the stainless steel mass in its weighing position. There are three strategies to mitigate this effect. First, a mass can be chosen that has a smaller permanent magnetization. As mentioned above, this calculation was performed with the maximum permissible magnetization for an E$_1$ mass. Second, the magnetization term changes sign as the mass is rotated up-side down. If a mass with a magnetization of  $|M_{\mathrm{z}}|\ge2$\,A/m has to be measured in the new watt balance, the magnetic effect can be nulled by averaging two measurements with the mass rotated up-side down in between. Third, two coils wired in series opposition  can be installed above and below the mass generating a magnetic field gradient. By choosing the right current in the coil a vertical field gradient can be generated that cancels the existing gradient. If the actual gradient at the mass is zero, both terms vanish.

In summary, the external field above the new permanent magnet is sufficiently small such that stainless steel masses can be used in the new watt balance. If an E$_1$ mass with a nominal value of 1\,kg is used, a worst case magnetic force of 4\,\mymu g is expected. 

\section{Power spectral amplitude of a coil inside the magnet}

\begin{figure}[h!]
\centering
\includegraphics[width=3.5in]{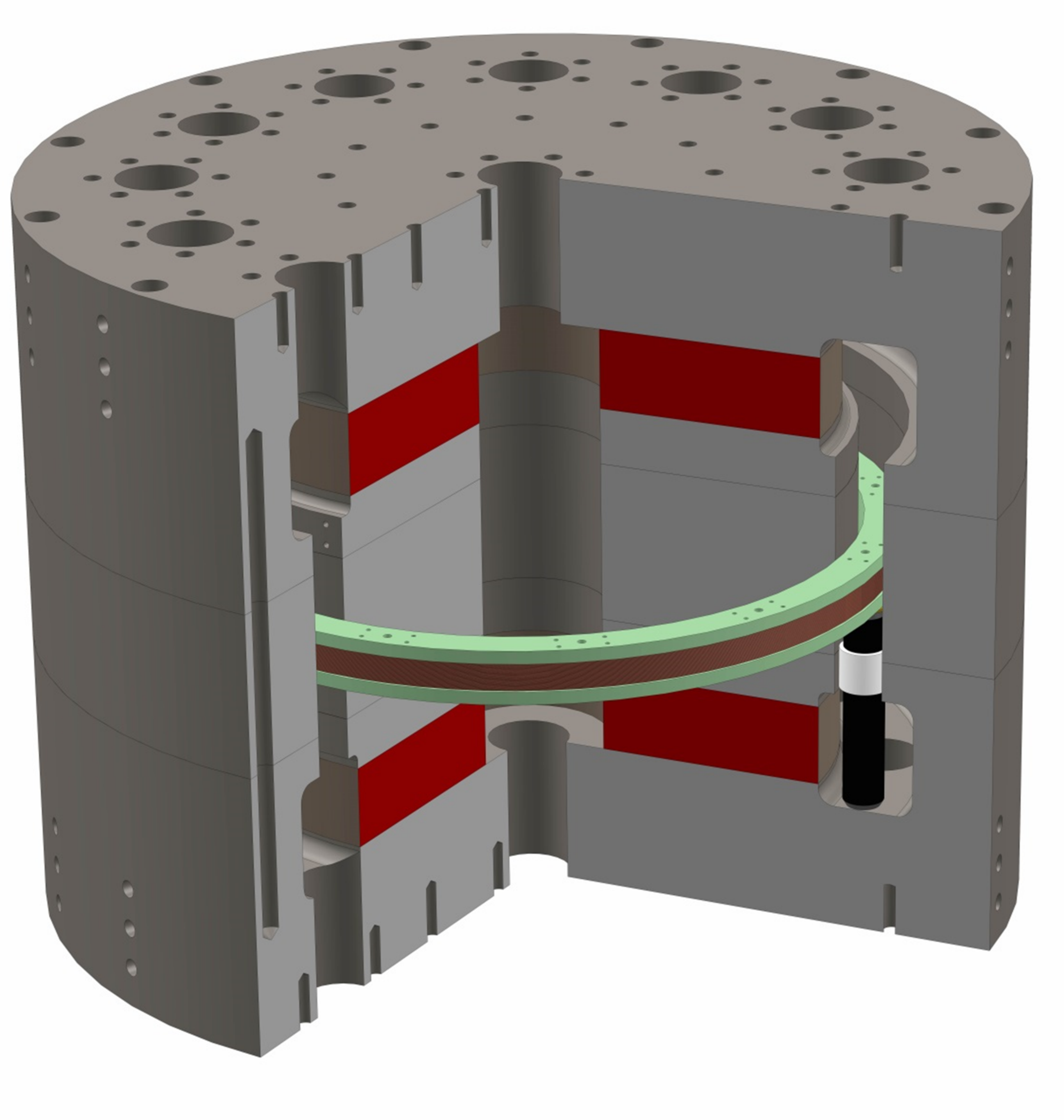}
\caption{The coil in the magnet. The coil is supported by three pillars, each comprised of  two optical posts and one Teflon spacer. Each pillar is attached to the bottom of the coil with brass set screws. The Teflon spacer centers the coil in the gap and eliminates horizontal motion.}
\label{fig:coilInGap}
\end{figure}

Another interesting measurement is the power spectral amplitude of the voltage across a coil at rest inside the magnet. To accomplish this measurement, the three coils of the radial gradiometer coil were connected in series. Fig.~\ref{fig:coilInGap} shows the coil in the gap. The coil is supported by three pillars. Each pillar is composed of  two optical posts and one Teflon spacer joined together by brass set screws. The coil is concentric and vertically centered inside the air gap. This measurement was performed with the magnet sitting on a pallet in storage. Hence, the vibrational environment for this measurement was not ideal. The power spectral amplitude was measured using a Rhode \& Schwarz UPV Audio analyzer. 

\begin{figure}[h!]
\centering
\includegraphics[width=3.5in]{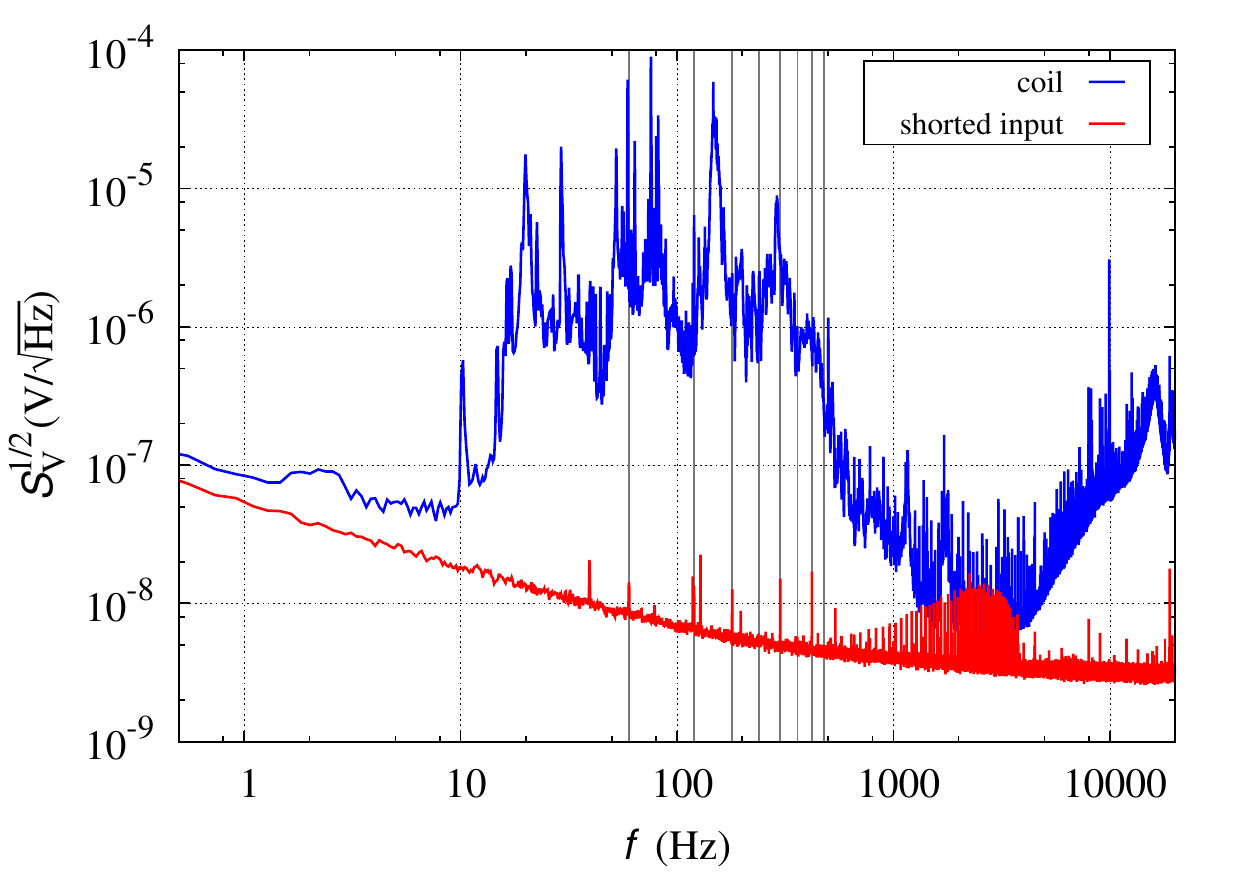}
\caption{Power spectral amplitude of terminal voltage of the radial gradiometer coil in the precision gap of the magnet. This measurement was performed with the magnet sitting on a pallet in storage. Hence the vibrational environment for this measurement was not ideal leading to the vibrational peaks in the spectral amplitude.}
\label{fig:frank2noise}
\end{figure}

Fig.~\ref{fig:frank2noise} shows the measured spectra for both channels of the analyzer. One channel was connected to the coil, while the other was shorted. Two observations on the spectrum of the coil voltage are noteworthy. First, at the low frequency end, the spectral amplitude is below 1\,\mymu V$\sqrt{\mbox{Hz}}$. This is an important figure of merit, since a white noise of smaller than 1\,\mymu V/$\sqrt{\mbox{Hz}}$ would allow the determination of the flux integral $Bl$ with a relative uncertainty of $10\times 10^{-9}$ in 10\,000 seconds. Second, there is a lot of excess noise in the region between 10\,Hz and 500\,Hz. This excess noise is mostly due to mechanical resonances in the coil and the coil support. These peaks are from different vibration modes of the coil, some of which we could identify. 

The radial gradiometer coil was not optimized for stiffness, since it was built to measure the radial gradient of the field. The design of the coil for NIST-4 is currently ongoing. One focus in the design work is to dampen the vibration mode in the frequency range from 10\,Hz to 300\,Hz. Ultimately, NIST-4 will be installled in an underground laboratory. This environment will have less vibration. In this environment the peaks in the spectral amplitude should be greatly reduced.

\section{Summary}
The NIST-4 magnet has been successfully built. Initial measurements of the basic properties of the magnet were carried out at NIST. A dedicated gradiometer coil was built to measure the vertical gradient of the radial flux density. The measurements with the gradiometer coil enabled the manufacturer, EEC, to regrind the gap improving the field profile. After delivery, it was found that the magnetic field profile could be further improved by changing the magnetic working point of the iron yoke. This can be accomplished by opening the magnet in a tilted fashion. Using this technique, the profile of the radial magnetic flux density could be changed to have a nearly vanishing derivative with respect to $z$ at the symmetry plane of the magnet. The magnetic flux density stayed within $\pm  10^{-4}$ of its value in the center over a travel range of 5\,cm. 

The radial dependence of the radial magnetic flux density was measured using a radial gradiometer coil. It was found that the field follows a $1/r$ dependence closely, and we expect any relative systematic error due to geometry changes of the coil of about $1.3\times 10^{-9}$.  This effect can be reduced by incorporating a heater in the coil.

We investigated the forces on the coil due to the reluctance force.  This force can lead to a systematic error via two mechanisms: (1) a difference in the mass-on and mass-off current and (2) a parasitic motion of the coil in these two states. It was determined that each of the two components produces a relative systematic error below $3\times 10^{-9}$ for reasonable assumptions.

The external magnetic field was measured above the magnet, i.e. where the balance and mass would be located. It was found that the field drops off rapidly reaching the earth's magnetic field at about 600\,mm above the top surface of the magnet. The spurious force on a stainless steel weight, class E$_1$, was calculated using worst case assumptions detailed in OIML R111. In this case, the relative systematic effect produced by the magnet is about $4\times 10^{-9}$. Hence, it is possible to use a E$_1$ stainless steel mass on the NIST-4 watt balance without a substantial increase in uncertainty. This uncertainty can be reduced by installing bucking coils, using PtIr artifacts or numerically canceling the permanent magnetization of the stainless artifact by measuring it upside down.

The power spectral amplitude of a coil in the magnet was measured. The spectrum is currently dominated by mechanical resonances and vibrations in the frequency region from 10\,Hz to 500\,Hz. Assuming these resonances can be removed and the vibrations damped, a measurement of $Bl$ with a relative uncertainty (type A) of  $10\times 10^{-9}$ can be achieved with an integration time of $\approx3$~hours or less. This uncertainty may be reduced due to partial cancellation of the voltage noise with velocity noise, which is likely to be highly correlated with the voltage noise.

Adding the relative type B uncertainties mentioned above in quadrature yields $\approx 5.2\times 10^{-9}$. This is a conservative estimate of the uncertainty of the magnet system, because the improvements outlined above would reduce this uncertainty by approximately a factor of two. In conclusion the known systematic effects from the magnet system are small enough to allow the construction of a watt balance at the 1\,kg level with a relative uncertainty of a few parts in $10^8$.


\begin{thebibliography}{1}

\bibitem{mills11}  I.M.~Mills, P.J.~Mohr, T.J.~Quinn, B.N.~Taylor and E.R.~Williams, "Adapting the International System of Units to the twenty-first century," {\it Phil. Trans. R. Soc A}., vol. 369, no. 1953, pp. 3907-3924, October, 2011.

\bibitem{steiner13} R. Steiner "History and progress on accurate measurements of the Planck constant," {\it Rep. Prog. Phys.}, vol. 76, no. 1, pp. 1-46, December, 2012. 

\bibitem{kibble75} Kibble B P, "A measurement of the gyromagnetic ratio of
the proton by the strong field method", {\it Atomic Masses and Fundamental Constants} vol. 5, ed J H Sanders and A H Wapstra (New York: Plenum), pp. 545-51, 1976.

\bibitem{Olsen80} P.T.~Olsen, W.D.~Phillips and E.R.~Williams, "A proposed coil system for the improved realization of the absolute Ampere", {\it J. Res. NBS}, vol. 85, pp. 257-72, July, 1980.

\bibitem{Schlamminger14} S.~Schlamminger, D.~Haddad, F.~Seifert, L.S.~Chao, D.B.~Newell, R.~Liu, R.L.~Steiner and J.R.~Pratt, "Determination of the Planck constant using a watt balance with a superconducting magnet system at the National Institute of Standards and Technology", {\it Metrologia}, vol. 51, pp. S15-S24, March, 2014.

\bibitem{ss13} S. Schlamminger, "Design of the permanent-magnet system for NIST-4," {\it IEEE Trans. Instrum. Meas.}, vol. 62, no. 6, pp. 1524-1530, June, 2013.

\bibitem{ms07} M. Stock, "Watt balances and the future of the kilogram," {\it INFOSIM Informative Bulletin of the Inter American Metrology System}, vol. 9, pp. 9-13, November, 2006.

\bibitem{Bozorth} Bozorth, R. M. {\it Ferromagnetism,} 1st ed New York, NY, USA: IEEE Press, 1993.

\bibitem{eichenberger04} A.L.~Eichenberger, J.~Butty, B.~Jeanneret, B.~Jeckelmann, A.~Joyet, T.~ Krebs, P.~Richard {\it et al}, "A new magnet design for the METAS watt Bblance," {\it Digest of the 2004 Conference on Precision Electromagnetic Measurements}, pp. 56-57, July, 2004.  

\bibitem{bnm} P. Gournay, G. Genev\`{e}s, F.~Alves, M.~Besbes, F.~Villar, and J.~David {\it et al}, "Magnetic circuit design for the BNM watt balance experiment," {\it IEEE Trans. Instrum. Meas.}, vol. 54, no. 2, pp.742-745, April, 2005.
 
\bibitem{Robinson12} I.A.~Robinson, "Towards the redefinition of the kilogram: a measurement of the Planck constant using the NPL Mark II watt balance", {\it Metrologia}, vol. 49, no. 1, pp. 113-16, December, 2011.

\bibitem{Gonzales65} G.R.~Gonzales and A. Brambilla, "Frequency dependence of the resistance and inductance of solid core magnets," {\it IEEE Trans. Nucl. Sci.}, vol. 12, no. 2, 349-353, June, 1965.

\bibitem{davis95} R.S. Davis, "Determining the magnetic properties of 1 kg mass standards," {\it J. Res. Nat. Inst. Std. Technol.}, vol. 100, no. 3, pp. 209-226, May, 1995.  

\bibitem{oiml111} Organisation Internationale de M'{e}trologie L'{e}gale, International, 
"International Recommendation on Weights of Classes E1, E2, F1, F2, M1, M2, M3," International Recommendation No. R111, 1994.

\end{thebibliography}
\end{document}